\documentclass[11pt]{article}
\usepackage[english]{babel}
\usepackage{amsmath,amsthm}
\usepackage{mathrsfs}
\usepackage{amsfonts}
\usepackage{graphicx}
\usepackage{enumerate}
\usepackage[usenames,dvipsnames]{color}
\usepackage{subfigure}
\usepackage[right,pagewise,displaymath, mathlines]{lineno}
\usepackage{epstopdf}
\usepackage{color}
\usepackage{multirow}
\usepackage[colorlinks = true,urlcolor = blue,citecolor = blue,breaklinks]{hyperref}
\usepackage{natbib}
\usepackage{dsfont}
\usepackage{amssymb}

\usepackage{pgfplots}
\usepackage{amssymb}
\usepackage{url}
\usepackage{fancyhdr}
\usepackage{indentfirst}
\usepackage{titlesec}
\usepackage{dsfont}
\usepackage[misc]{ifsym}

\def\d{\,\mathrm{d}}
\def\laweq{\buildrel \mathrm{d} \over =}

\newcommand{\VaR}{\mathrm{VaR}}

\newcommand{\ES}{\mathrm{ES}}
\newcommand{\E}{\mathbb{E}}
\newcommand{\R}{\mathbb{R}}

\newcommand{\mM}{\mathcal{M}}

\newcommand{\N}{\mathbb{N}}
\newcommand{\p}{\mathbb{P}}

\newcommand{\id}{\mathds{1}}

\newcommand{\X}{\mathcal X}

\newcommand{\esssup}{\mathrm{ess\mbox{-}sup}}

\renewcommand{\ge}{\geqslant}
\renewcommand{\le}{\leqslant}
\renewcommand{\geq}{\geqslant}
\renewcommand{\leq}{\leqslant}
\renewcommand{\epsilon}{\varepsilon}

\theoremstyle{plain}
\newtheorem{theorem}{Theorem}
\newtheorem{lemma}{Lemma}
\newtheorem{proposition}{Proposition}

\theoremstyle{definition}

\newtheorem{example}{Example}[section]

\theoremstyle{remark}
\newtheorem{remark}{Remark}

\theoremstyle{definition}

\numberwithin{equation}{section}
\numberwithin{theorem}{section}
\numberwithin{proposition}{section}
\numberwithin{lemma}{section}
\numberwithin{remark}{section}
\renewcommand{\cite}{\citet}

\usepackage{geometry}
\usepackage[onehalfspacing]{setspace}

\begin{document}

\title{Cash-subadditive risk measures without quasi-convexity}

\author{Xia Han\thanks{School of Mathematical Sciences and LPMC, Nankai University, China. Email: xiahan@nankai.edu.cn} \and Qiuqi Wang\thanks{ Maurice R.~Greenberg School of Risk Science, Georgia State University, U.S.A. Email: qwang30@gsu.edu}\and Ruodu Wang\thanks{Department of Statistics and Actuarial Science, University of Waterloo, Canada. Email: wang@uwaterloo.ca}\and Jianming Xia\thanks{Key Laboratory of Random Complex Structures and Data Science (RCSDS), National Center for Mathematics and Interdisciplinary Sciences (NCMIS), Academy of Mathematics and Systems Science, Chinese Academy of Sciences, China. Email: xia@amss.ac.cn.}}

\date{\today}

\maketitle

\begin{abstract} 
In the literature {on} risk measures, 
cash subadditivity was proposed to replace cash additivity, motivated by the presence of stochastic or ambiguous interest rates and defaultable contingent claims. Cash subadditivity has been traditionally studied together with quasi-convexity, in a way similar to  cash additivity with convexity. 
  In this paper, we study cash-subadditive risk measures without quasi-convexity.
One of our major results is that a general cash-subadditive risk measure  can be represented as the lower envelope of a family of quasi-convex and cash-subadditive risk measures.
Representation results of cash-subadditive risk measures with some additional properties are also examined.  The notion of  quasi-star-shapedness, which is a natural analogue of  star-shapedness, is introduced, and  we obtain a corresponding representation result via the lower envelope of normalized,  quasi-convex and cash-subadditive  risk measures.

\medskip
\noindent\emph{
}\textsc{Keywords:}  cash subadditivity, quasi-convexity, stochastic dominance, star-shapedness, Lambda-VaR
\end{abstract}

\section{Introduction}
The quantification of market risk for pricing, portfolio selection, and risk management purposes has long been a point of interest to researchers and practitioners in finance. Measures of risk have been widely adopted to assess the riskiness of financial positions and determine capital reserves.
Value-at-risk (VaR) has been one of the most commonly adopted risk measures in industry but is criticized due to its fundamental deficiencies; for instance, it does not account for ``tail risk"  and  it lacks for subadditivity or convexity; see e.g.,  \cite{DEGKMRS01} and \cite{MFE15}.  
In light of this, the notion of  coherent risk measures  that satisfy a set of reasonable axioms (monotonicity, cash additivity, subadditivity and positive homogeneity) was introduced by  \cite{ADEH99} and extensively treated by \cite{D02}.  
Convex risk measures were introduced by \cite{FR02} and \cite{FS02a} with convexity replacing subadditivity and positive homogeneity.  There have been many other developments in the past two decades in various directions; see \cite{FS16} and the references therein.
  
  A common feature of all above risk measures is that  the axiom of  cash additivity (also called cash invariance or translation invariance)  is employed. 
The cash additivity axiom has been challenged, in particular, by  \cite{KR09}, in a relevant context. The  main motivation for   cash additivity is that the random losses should be discounted by a constant num\'eraire.
 Therefore,  cash additivity fails   as soon as there is any form of uncertainty about interest rates.  For this reason,  \cite{KR09} replaced cash additivity by cash subadditivity and provided a representation result for convex cash-subadditive risk measures. In this context,  \cite{CMMM11} argued that quasi-convexity rather than convexity is the appropriate mathematical translation of the statement ``diversification should not increase the risk" and introduced the notion of  quasi-convex cash-subadditivie  risk measures.   \cite{FKM14}  studied general risk measures  to model defaultable contingent claims and discussed their relationship with cash-subadditive risk measures, and  {\cite{AM21} further  studied risk measures beyond frictionless markets.}  For other  related  work on cash subadditivity and quasi-convexity, see \cite{FM11}, \cite{CDH13}, \cite{DK13}, \cite{FMP14} and \cite{M15}. In decision theory, the economic counterpart of quasi-convexity of risk measures is quasi-concavity of utility functions, which is classically associated to uncertainty aversion in the economics of uncertainty; see, e.g., \cite{S89}, \cite{CMMM11}, \cite{MR15} and \cite{LS19}. 
 
 The main aim of this paper is a   thorough understanding of cash-subadditive risk measures when quasi-convexity, or the stronger property of convexity, is absent.  
This class of risk measures is very broad and, with proper normalization, it contains a wide majority of risk measure or preference functional {considered} in the literature. By relaxing cash additivity,  both the theory of risk measures and that of expected utility and rank-dependent utility (\cite{Q82}) can be included within the same framework.  For instance, the mapping $X\mapsto -\E[u(-X)]/m$ for  any increasing utility function $u$ with derivative bounded above by $ m>0$ belongs to this class (recall that the constant $m$ does not matter when modeling utility preferences); the same holds true if $\E$ is replaced by a non-additive and normalized Choquet integral (\cite{Y87, S89}). Here, {the  utility function $u$} may not be convex or concave; see the recent discussions and examples in  \cite{MSTW17} and \cite{CCMTW21} for non-convex and non-concave loss and utility functions.
 
Cash-additive risk measures without convexity have been actively studied  in the recent literature.  In particular, a few representation results were obtained by \cite{MW20}, \cite{JXZ21} and \cite{CCMTW21}.  As a common feature, such risk measures can be represented as the infimum over a collection of convex and cash-additive risk measures (see Table \ref{tab:1} below),\footnote{The  risk measures  studied  by \cite{JXZ21} and \cite{CCMTW21} have similar representations; their differences are studied by \cite{MR22}.}  in contrast to the  classic theory of convex risk measures where representations are typically based on a supremum.  %
In a similar fashion, one of our main results states that a general cash-subadditive risk measure can be represented as the lower envelope of a family of quasi-convex cash-subadditive risk measures.

In addition to a representation of general cash-subadditive risk measures, we will also give implicit and explicit representations of cash-subadditive risk measures with additional properties including quasi-star-shapedness,  normalization {(that is, $\rho(x)=x$ for all real $x$)}  and SSD-consistency (that is, consistency with second-order stochastic
dominance).  In particular, similarly to the argument that convexity does not fit well with cash-subadditive risk measures, star-shapedness introduced by \cite{CCMTW21} is no longer a natural property beyond the framework of cash-additive risk measures.   In this sense, we introduce the property of quasi-star-shapedness induced naturally from quasi-convexity, and obtain a representation result of cash-subadditive risk measures that are normalized and quasi-star-shaped. It turns out that the representation result also holds true if we change normalization to a weaker version which we call quasi-normalization.
 

We examine a few other problems studied by  \cite{MW20}, now under a general framework of cash subadditivity. Apart from the major differences, it also turns out that some of   results obtained by \cite{MW20}   hold under the extended framework of cash subadditivity.  A comparison of our results and some results in the literature is summarized in Section \ref{sec:6}.

The new property of quasi-star-shapedness 
 has a sound decision-theoretic foundation. Translating it into the setting of \cite{AA63}, it means that the decision maker always prefers to replace part of an uncertain (random) payoff with an equally favourable certain (non-random) payoff. This property is a weaker requirement than the uncertainty aversion axiom studied by \cite{MMR06}, which corresponds to quasi-convexity in our setting.

The rest of the  paper is organized as follows. In Section \ref{sec:pre},  some preliminaries on risk measures are collected and the definition of   cash-subadditive risk measures is given.
Two new properties, quasi-star-shapedness and quasi-normalization, are introduced in Section \ref{sec:new3} and we provide a few  related results. In particular, we obtain  a new formula (Theorem \ref{prop:lambdarep}) for $\Lambda\VaR$ introduced by \cite{FMP14}, which is an example of quasi-star-shaped, quasi-normalized   and cash-subadditive risk measures. {Quasi-star-shapedness of other cash-subadditive risk measures is also discussed.}
In Section \ref{sec:3}, 
representation results for general cash-subadditive risk measures are established.
Section \ref{sec:extend} contains representation results and other technical results on cash-subadditive risk measures with further properties including quasi-star-shapedness and SSD-consistency. 
Section \ref{sec:6} concludes the paper, and the appendix contains some further technical results and discussions that are not directly used in the main text. 
As the main message of this paper, most of the existing results on non-convex cash-additive risk measures have a nice parallel version {for} non-quasi-convex cash-subadditive risk measures, although they often require more sophisticated analysis to establish.

\section{Cash-subadditive risk measures}\label{sec:pre}
Let $(\Omega, \mathcal{F},P)$ be a probability space, 
$
\mM_f
$
be the set of  finitely additive probabilities on $(\Omega, \mathcal{F})$ that are absolutely continuous with respect to $P$, 
and 
$\mM$ represent the subset of $\mM_f$ consisting of all its countably additive elements, i.e., probability measures.  Let $\X=L^\infty(\Omega, \mathcal{F},P)$ be the set of all essentially bounded random variables on $(\Omega, \mathcal{F}, P)$, where $P$-a.s.~equal random variables are treated as identical.\footnote{As such, equalities and inequalities   should be understood in a $P$-a.s.~sense.}
Let a random variable $X\in \X$ represent the random loss faced by financial institutions in a fixed period of time; {  a positive
value of $X\in \X$  represents a loss and a negative $X$ represents a surplus; this sign convention is used by, e.g., \cite{MFE15}.}
We write $X \laweq Y$
if two random variables $X,Y\in \X$ follow the same  distribution   under $P$. {Throughout the paper,  ``increasing" and ``decreasing" are in the  nonstrict {(weak)} sense},   
$a\vee b$  (resp.~$a\wedge b$) is the maximum (resp.~minimum) between real numbers $a$ and $b$, and $a_+=a\vee 0$.

A mapping $\rho: \X \rightarrow \R$ is called a \emph{risk measure} if it satisfies:
\begin{itemize}
 \item[] \emph{Monotonicity}: $\rho(X) \leq \rho(Y)$ for all $X,Y\in \X$ with $X \leq Y$.
 \end{itemize}
Monotonicity is self-explanatory and common in the literature {on} risk management, e.g., \cite{ADEH99}.  It means that if the loss increases for almost all scenarios $\omega\in\Omega$, then the capital requirement in order for the financial position to be acceptable should increase as well. The risk measure $\rho$ is called a \emph{monetary risk measure} if it further satisfies
\begin{itemize}
\item[] \emph{Cash additivity}: $\rho(X+m)=\rho(X)+m$ for all $X\in \X$ and $m\in\R$.
\end{itemize}
Cash additivity (also called cash invariance or translation invariance) intuitively means that the risk measure $\rho(X)$ is the amount of capital that needs to be added to the financial position $X$ to make it acceptable.
Cash additivity is a nice and simplifying mathematical property, but the class of cash-additive risk measures is too restricted to include some common functionals such as the expectation of a convex loss function. From the viewpoint of financial practice, the assumption of cash additivity of a risk measure may   fail when uncertainty of interested rates is taken into account. In this sense, we consider the more general class of risk measures $\rho$, as argued by \cite{KR09}, satisfying
\begin{itemize}
\item[] \emph{Cash subadditivity}: $\rho(X+m)\leq\rho(X)+m$ for all $X\in \X$ and $m\ge 0$.\footnote{An equivalent definition of cash subadditivity is  $\rho(X+m)\geq\rho(X)+m$ for all $X\in \X$ and $m\le 0$.}
\end{itemize}
The assumption of cash subadditivity  allows non-linear increase of the capital requirement as cash is added to the financial position but the increase should not exceed linear growth.  {Moreover, for a mapping $\rho:\X\to (-\infty,\infty]$,  cash subadditivity of $\rho$ implies that $\rho$ is finite everywhere as soon as  it is finite somewhere, and hence we can focus only on real-valued mappings.}
 \begin{remark}\label{rem:continuity} Cash-subadditive risk measures are $L^\infty$-continuous;  namely, $\lim _{n \rightarrow \infty} \rho\left(X_{n}\right)=\rho(X)$ for any sequence $X_{n} \in \mathcal{X}$ satisfying $\esssup \left(\left|X_{n}-X\right|\right) \rightarrow 0$ as
$n \rightarrow \infty$.  Clearly, for all $X, Y \in \mathcal{X}$,  $X \leq Y+\|X-Y\|$.  By monotonicity and cash subadditivity of $\rho$, we have   $\rho(X)-\rho(Y) \leq \|X-Y\|$. Switching the roles of $X$ and $Y$ yields that $\rho$ is $1$-Lipschitz $L^\infty$-continuous.
Conversely, if a risk measure is $1$-Lipschitz $L^\infty$-continuous,
then $\rho(X+m)-\rho(X) \le   \|m\| =m$ for all $m\ge 0$. Therefore, 
$1$-Lipschitz $L^\infty$-continuity and cash subadditivity are equivalent for risk measures; this observation is made in Proposition 2.1 of \cite{CMMM11}. 
\end{remark}
Cash-subadditive risk measures are often studied in the literature together with convexity, or more generally, with quasi-convexity; see \cite{KR09}, \cite{CMMM11} and \cite{FMP14}.
\begin{itemize}
\item[] \emph{Convexity}: $\rho(\lambda X+(1-\lambda) Y) \leq \lambda \rho(X)+(1-\lambda) \rho(Y)$ for all $X,Y\in \X$ and  $\lambda \in[0,1]$.
\item[] \emph{Quasi-convexity}: $\rho(\lambda X+(1-\lambda) Y) \leq \max \{\rho(X), \rho(Y)\}$ for all $X,Y\in \X$ and $\lambda \in[0,1]$.
\end{itemize}

As the main objective of this paper is to study cash-subadditive risk measures without quasi-convexity, 
we first note that the lack of quasi-convexity arises in many economically relevant contexts, such as aggregation of risk measures, non-convex utility functions, and risk mitigation.
\cite{CCMTW21} argued with examples that many operations on a collection of convex risk measures lead to a non-convex one; the same applies in the context of quasi-convexity.
Other than those built from operations, we provide a few simple examples of cash-subadditive risk measures  in the literature, which are not cash-additive or quasi-convex.
 
 
 The risk measure  \emph{Value-at-Risk} (VaR) is
given by, for $t\in (0,1]$, 
\begin{equation}\label{eq:vardef}
\VaR_t(X)=\inf\{x\in\R\mid P(X\le x)\ge t\},~~~X\in \X.\end{equation}
Note that $\VaR_1(X)=\esssup(X)$.
  VaR is one of the most popular risk measures used in the banking industry; see \cite{MFE15}.  The next example is a  generalization of VaR  introduced by \cite{FMP14} without cash additivity. 
\begin{example}[$\Lambda$-Value-at-Risk]
\label{exm:5} 
The risk measure $\Lambda$-Value-at-Risk is defined as,
for some function  $\Lambda: \R\to[0,1]$ that is not constantly $0$,
\begin{equation}\label{eq:lambdadef}
\Lambda\VaR(X) = \inf \{x\in \R: P(X\le x) \ge \Lambda(x)\},~~~X\in \X.
\end{equation}
In particular, if $\Lambda$ is a constant  $t\in (0,1)$, then $\Lambda\VaR = \VaR_t$.
{Although \cite{FMP14} mainly studied   increasing $\Lambda$,
the recent work of \cite{BPR17} and \cite{BP21} has shown that using a decreasing $\Lambda$ leads to many advantages, including, robustness, elicitability, and an axiomatic characterization. For this reason, we assume that  $\Lambda$ is  a decreasing function in this paper.}
Since for $c\ge 0$, 
  $\Lambda\VaR(X+c) = \Lambda^*\VaR(X)+c $ where $\Lambda^*(t)=\Lambda(t+c)\le \Lambda(t)$ for $t\in \R$, 
we obtain $\Lambda\VaR (X+c) \le \Lambda\VaR(X)+c$, and therefore $\Lambda\VaR$ is cash subadditive; we can check that it is not cash additive in general.\footnote{ {Note that $\Lambda\VaR(X)$ is not necessarily cash subadditive when $\Lambda$ is increasing.  For instance, take $\Lambda: x\mapsto  (3/4)\id_{\{x > 1/2\}} +(1/2)\id_{\{x\le 1/2\}} $. 
Let $X$ be uniformly distributed on $[0,1]$ and take $\epsilon>0$.
We have $\Lambda \VaR(X)=1/2$ and $\Lambda(X+\epsilon)=3/4+\epsilon$,
and this clearly violates cash subadditivity. Moreover, such $\Lambda\VaR$ is not even continuous with respect to $L^\infty$-norm, which is a basic requirement for risk measures interpreted as capital reserve.}}
Moreover,  $\Lambda\VaR$ is generally not quasi-convex either, as the following argument illustrates. 
For any   decreasing $\Lambda:\R\to (0,1/3]$ and 
a standard normal random variable $X$, 
we have $  \Lambda\VaR(X)=\Lambda\VaR(-X)\le z_{1/3} <0$,
where $z_{1/3}$ is the $1/3$-quantile of the standard normal distribution.  
Hence, $\Lambda\VaR(0)= 0 > \max \{ \Lambda\VaR(X), \Lambda\VaR(-X)\}$ violating quasi-convexity.
\end{example}

 \begin{example}[Expected insured loss]\label{ex:put}
Suppose that an insurance contract pays $f(X)$ for an insurable loss $X$ (often non-negative), where $f$ is an increasing function on $\R$ that is $1$-Lipschitz and $f(x)=0$ for $x\le 0$.\footnote{A function $f:\R\to\R$ is called \emph{$1$-Lipschitz} if $|f(x)-f(y)|\le|x-y|$ for all $x,y\in\R$.} 
A typical example is $f(x)= (x-d)_+ \wedge \ell$ for some $\ell>d>0$, which represents an insurance contract with deductible $d$ and limit $\ell$.
The   expected losses to the policy holder and
to the insurer are 
 given by, respectively,
 $$
 \rho_{\rm ph}(X) = \E[X-f(X)] \mbox{~~~and~~~}  \rho_{\rm in} (X)=\E[f(X)].
 $$ 
It is straightforward to check that $\rho_{\rm ph}$ and $\rho_{\rm in}$ are both monotone and cash subadditive, but generally neither cash additive nor quasi-convex. In particular, $\rho_{\rm in}$ (resp.~$\rho_{\rm ph}$) is concave if $f$ is concave (resp.~convex). 
For a related example in finance, take $f:x\mapsto x_+  $ and fix a probability measure $Q$  representing a pricing measure in a financial market.  The put option premium on the insolvency of a firm with future asset value $-X$ is defined as $\E_Q[X_+]$, which is convex and cash subadditive but not cash additive; see \cite{J02} and \cite{KR09} for a connection between the put option premium and risk measures.
 \end{example}

 \begin{example}[Certainty equivalent with discount factor ambiguity]\label{ex:alphaMEU}
Consider the following $\alpha$-maxmin expected utility ($\alpha$-MEU, \cite{M02, GMM04}) with a profit-loss adjustment: 
\begin{equation*}
\alpha\min_{Q_1\in\mathcal{Q}_1}\E_{Q_1}[\mathrm{e}^{\gamma X}]+(1-\alpha)\max_{Q_2\in\mathcal{Q}_2}\E_{Q_2}[\mathrm{e}^{\gamma X}],~~X\in\X,~\alpha\in[0,1],~\gamma>0
\end{equation*}
with the loss function $x\mapsto \mathrm{e}^{\gamma x}$, where $\mathcal{Q}_1$ and $\mathcal{Q}_2$ are two nonempty, weak*-compact and convex sets of finitely additive probabilities. 
{The $\alpha$-MEU model is of   interest to the applied decision-theoretic literature as it can model deviations from the pure pessimism that the MEU model (corresponding to $\alpha=0$) inherently embodies. The suggestion that decision makers are not purely pessimistic is supported by a vast amount of empirical literature,  see e.g., \cite{TV15} for a review.}

{The certainty equivalent of the above $\alpha$-MEU with  a  stochastic ambiguous discount factor  $D$ is given by
$$\rho(X)=\sup_{D\in \mathcal I}\left\{\frac{1}{\gamma}\log\left(\alpha\min_{Q_1\in\mathcal{Q}_1}\E_{Q_1}[\mathrm{e}^{\gamma D X}]+(1-\alpha)\max_{Q_2\in\mathcal{Q}_2}\E_{Q_2}[\mathrm{e}^{\gamma D X}]\right)\right\},~~X\in\X,$$
where $\mathcal I$ is a set of random variables taking values in $[0,1]$. 
For any $X\in\X$ and  $m\geq 0$,   we have 
\begin{align*}\rho(X+m)&=\sup_{D\in \mathcal I}\left\{\frac{1}{\gamma}\log\left(\alpha\min_{Q_1\in\mathcal{Q}_1}\E_{Q_1}[\mathrm{e}^{\gamma D (X+m)}]+(1-\alpha)\max_{Q_2\in\mathcal{Q}_2}\E_{Q_2}[\mathrm{e}^{\gamma D( X+m)}]\right)\right\}\\&\leq \sup_{D\in \mathcal I}\left\{\frac{1}{\gamma}\log\left(\alpha e^{\gamma m}\min_{Q_1\in\mathcal{Q}_1}\E_{Q_1}[\mathrm{e}^{\gamma D X}]+(1-\alpha)e^{\gamma m}\max_{Q_2\in\mathcal{Q}_2}\E_{Q_2}[\mathrm{e}^{\gamma D X}]\right)\right\}\\&= \sup_{D\in \mathcal I}\left\{\frac{1}{\gamma}\log\left(\alpha  \min_{Q_1\in\mathcal{Q}_1}\E_{Q_1}[\mathrm{e}^{\gamma D X}]+(1-\alpha)\max_{Q_2\in\mathcal{Q}_2}\E_{Q_2}[\mathrm{e}^{\gamma D X}]\right)\right\}+m= \rho(X)+m. \end{align*}
Therefore,   $\rho$ is a cash-subadditive risk measure, and  it is generally not cash additive.}  Moreover, because of the presence of both minimum and maximum in   $\alpha$-MEU, quasi-convexity does not hold for $\rho$.
\end{example}

\begin{example}[Risk measures based on    eligible risky assets]
\label{exm:4}
Take an acceptance set $\mathcal{A} \subseteq \X$ and a reference asset $S=\left(S_{0}, S_{T}\right)\in\X^2$, where  $S_0$ is a constant representing the initial price of the asset, and
$S_{T}$ is a positive terminal payoff.  Define the mapping $\rho_{\mathcal{A}, S}$ as in \cite{FKM14}  by
\begin{equation}\label{eq:elig}\rho_{\mathcal{A}, S}(X)=\inf \left\{m \in \mathbb{R}: X-\frac{m}{S_{0}} S_{T} \in \mathcal{A}\right\},~~X\in\X.\end{equation}
The quantity $\rho_{\mathcal{A}, S}(X)$ represents the ``minimal" amount of capital we have to raise and invest, at inception, in the asset $S$ to meet the acceptability constraint specified by $\mathcal{A}$. 
By Proposition 5.1 of \cite{FKM14}, we have that $\rho_{\mathcal{A}, S}$ is cash subadditive under the assumption of $P\left(S_T<S_0\right)=0$ (e.g., the bond can only default on the interest payment).\footnote{Beyond the sufficiency of the condition $\mathbb{P}\left(S_T<S_0\right)=0$, for the acceptance set $\mathcal{A}$ chosen based on the common risk measures ES or VaR, if $\rho_{\mathcal{A}, S}$ is cash subadditive, then  $\mathbb{P}\left(S_T<S_0\right)$ needs to be sufficiently small; see Corollary 5.3 and Proposition 5.5 of \cite{FKM14}. In particular, $P\left(S_T<S_0\right)=0$ is    a necessary condition for cash subadditivity with a VaR-based acceptance set.}  In this case, assuming that $\mathcal{A}$ is closed, $\rho_{\mathcal{A}, S}$ is convex if and only if $\mathcal{A}$ is convex (see Lemma 2.5 of \cite{FKM14}).\footnote{When the asset is not liquidly traded,  we can define  $\rho_{\mathcal{A}, S,\pi}(X)=\inf \left\{\pi(m) \in \mathbb{R}: X-m S_{T} \in \mathcal{A}\right\}, X\in\X,$ where $\pi:\mathbb R\rightarrow\mathbb R$ is some non-linear increasing function. Then  $\rho_{\mathcal{A}, S, \pi}$ is quasiconvex if and only if $\mathcal A$ is convex because the composition of convex and increasing functions leads to $\rho_{\mathcal{A}, S}$ being quasi-convex.}   In general, such risk measures are not quasi-convex {unless they are also convex}. 

{ Although $\rho_{\mathcal A,S}$ is not cash additive, by definition it is affine along $S$, that is,
 $
\rho_{\mathcal A,S} (X+\lambda S_T) = \rho_{\mathcal A,S}(X) + \lambda S_0
 $ for all $\lambda \in \R$ and $X\in \mathcal X$. Conversely, we can check that
 all risk measures that are affine along $S$ can be identified as $\rho_{\mathcal A,S}$, where $\mathcal A=\{X\in \X: \rho(X)\le 0\}$. Therefore, all results on cash-subadditive risk measures apply to such risk measures as long as $P(S_T<S_0)=0$.}

 To account for more than one eligible asset, we consider two assets for an example.  Fix two acceptance sets $\mathcal{A}_1$ and $\mathcal{A}_2$ in $\mathcal{X}$, and two assets $S^1=\left(S_0^1, S_T^1\right)$ and $S^2=\left(S_0^2, S_T^2\right)$. Moreover,  define  the set
$$
\mathcal{P}_0\left(S^1, S^2\right):=\left\{\frac{m}{S_0^2} S_T^2-\frac{m}{S_0^1} S_T^1 ; ~ m\in \mathbb{R}\right\}.\footnote{This set consists of the payoffs of all ``portfolios" we can form at zero cost by combining the assets $S^1$ and $S^2$.}
$$  
  By Proposition 3.2.12 of  \cite{M15}, the inf-convolution of $\rho_{\mathcal{A}_1, S^1}$ and $\rho_{\mathcal{A}_2, S^2}$ has the following representation
 $$\rho_{\mathcal{A}_1, S^1} \square \rho_{\mathcal{A}_2, 
S^2}=\rho_{\mathcal{A}_1+\mathcal{A}_2+\mathcal{P}_0\left(S^1, S^2\right), S^1}.\footnote{For a fixed position $X \in \mathcal{X}$, the inf-convolution of  $f_1: \mathcal{X} \rightarrow \mathbb{R} \cup\{\infty\}$ and $f_2: \mathcal{X} \rightarrow \mathbb{R} \cup\{\infty\}$ is the map $f_1 \square f_2: \mathcal{X} \rightarrow {\mathbb{R}} \cup\{\infty\}$ defined by
$
f_1 \square f_2(X):=\inf \left\{f_1(Y)+f_2(X-Y) ; Y \in \mathcal{X}\right\}. 
$ As such, $f_1 \square f_2(X)$ is the ``minimal" total required capital across all possible allocations of the aggregated position $X$.}$$  
Thus,  $\rho_{\mathcal{A}_1, S^1} \square \rho_{\mathcal{A}_2, 
S^2}$ is still cash subadditive; this can certainly be extended to a finite number of assets.
For more details on inf-convolutions and their applications to the theory of
 risk measures, we refer to \cite{BE05} and \cite{FS08}. 
\end{example}


Some other relevant properties  for a risk measure $\rho$ are collected below, which will be used throughout the paper; we refer to \cite{FS16} for a comprehensive treatment of properties of risk measures. 
\begin{itemize}
\item[] \emph{Normalization}: $\rho(t)=t$ for all $t\in\R$.
\item[] \emph{Law invariance}: $\rho(X)=\rho(Y)$ for all $X,Y\in \X$ with $X \laweq Y$.
\end{itemize}
{ Normalization here is more general than the  traditional definition of $\rho(0)=0$ in \cite{FS16}, meaning that  the risk of  any constant   equals itself. The two definitions are equivalent if $\rho$ is cash additive.}  Monetary, convex and positively homogeneous  risk measures are called coherent by \cite{ADEH99}.\footnote{The functional $\rho$ is said to be positively homogeneous if $\rho(\lambda X)=\lambda \rho(X)$ for all $X\in \X$ and $\lambda\ge 0$.}

Next, we define the two most important notions of stochastic dominance in decision theory,  the first-order stochastic dominance (FSD) and the second-order stochastic dominance (SSD). 
Given two random variables $X, Y \in \X $, we denote by $X \succeq_{1} Y$,  if
 $
\E [f(X)] \geq \E [f(Y)]
 $
for all increasing  functions $f:\R\to\R$, and denote by $X  \succeq_{2}  Y$, if $\mathbb{E}[f(X)] \geqslant \mathbb{E}[f(Y)]$ for all increasing convex functions $f: \mathbb{R} \rightarrow \mathbb{R}$. Consistency with respect to FSD or SSD is defined as monotonicity in these   partial orders.
\begin{itemize}
\item[] \emph{FSD-consistency}: $\rho(X) \geq \rho(Y)$ for all $X,Y\in \X$ whenever $X \succeq_{1} Y$.
\item[] \emph{SSD-consistency}: $\rho(X) \geq \rho(Y)$ for all $X,Y\in \X$ whenever $X \succeq_{2} Y$.   
\end{itemize}
It is well known that  either FSD-consistency or SSD-consistency implies  law invariance.
For monetary risk measures, SSD-consistency is characterized by Theorem 3.1 of \cite{MW20}. 

Finally, the notion of comonotonicity is useful for some results in this paper. A random vector $(X_1,\dots,X_n) \in \X^n$ is called \emph{comonotonic} if there exists a random variable $Z\in\X$ and increasing functions $f_1,\dots,f_n$ on $\R$ such that $X_i=f_i(Z)$ almost surely for all $i=1,\dots,n$.


\section{Quasi-star-shapedness, quasi-normalization, and Lambda VaR}
\label{sec:new3}

In this section, we discuss two new properties that are specific to cash-subadditive risk measures without quasi-convexity, and they will be used in the representation results in Section \ref{sec:PH}. 

\subsection{Quasi-star-shapedness and quasi-normalization}
In the context of cash-additive risk measures, \cite{CCMTW21} studied a weaker property than convexity:
\begin{itemize}
\item[] \emph{Star-shapedness}: $\rho(\lambda X)\le\lambda\rho(X) + (1-\lambda)\rho(0)$ for all $X\in \X$   and $\lambda\in[0, 1]$,
\end{itemize}    and formulated  star-shapedness via $\rho(\lambda X)\le\lambda\rho(X) $  for $\lambda \in [0,1]$ with the extra normalization   $\rho(0)=0$.  Star-shapedness  is discussed  in \cite{ADEH99} and it  has a natural economic motivation that  additional liquidity risk may arise if a position is multiplied by  a factor larger than $1$.  In case  $\rho(0)\ne 0$,    it is more natural to define star-shapedness via our formulation, which means  convexity at $0$ (has also been called “positive superhomogeneity” for obvious mathematical reasons),  thus weaker than convexity. 
In the context of the cash-additive risk measures,  we introduce the corresponding property for cash-subadditive risk measures:
\begin{itemize}
\item[] \emph{Quasi-star-shapedness}: $\rho(\lambda X+(1-\lambda)t)\le \max\{\rho(X),\rho(t)\}$ for all $X\in \X$, $t\in \R$  and $\lambda\in[0, 1]$. 
\end{itemize}

Since quasi-star-shapedness is new to the literature {on} risk measures, it may need some explanation.  As explained by \cite{CCMTW21}, star-shapedness reflects the consideration of liquidity risk, in a way similar to (but weaker than) convexity which reflects the consideration of diversification.  For  cash-additive risk measures, star-shapedness  is equivalent to
$\rho(\lambda X + (1-\lambda)t )\le\lambda\rho(X) +(1-\lambda)\rho(t)$ for all $X\in \X$, $t\in \R$  and $\lambda\in[0, 1]$; indeed, it means that $\rho$ has convexity at each constant. 
 This reformulation of star-shapedness implies our quasi-star-shapedness, which means  that $\rho$ has quasi-convexity at each constant. Obviously, quasi-star-shapedness is weaker than quasi-convexity.

Quasi-star-shapedness has a sound decision-theoretic interpretation, which we explain in Proposition \ref{prop:oct-1} below. 
For a risk measure $\rho: \X\to \mathbb{R}$, the preference associated with $\rho$ is a binary relation $\succeq$ on 
$\X$ defined by,  for all $X,Y\in \X$,  $X \succeq Y \Longleftrightarrow \rho(X)\le \rho(Y)$. The equivalence relation of $\succeq$ is denoted by $\simeq$.
In other words, $\succeq$ represents the preference of an agent favouring less risk evaluated via $\rho$.

\begin{proposition}\label{prop:oct-1}
An $L^\infty$-continuous risk measure  $\rho: \X\to \mathbb{R}$ satisfies quasi-star-shapedness
if and only if its associated preference $\succeq$ satisfies, for $X\in \X$, $t\in \R$ and  $\lambda\in [0,1]$, 
\begin{align}\label{eq:oct-1}
X\simeq t ~\Longrightarrow~  \lambda X+(1-\lambda)t  \succeq X.
\end{align}
\end{proposition}
\begin{proof}
By definition of $\succeq$, \eqref{eq:oct-1} is equivalent to 
\begin{align}\label{eq:oct-2}
\rho(X)=\rho(t) ~\Longrightarrow~  \rho(\lambda X+(1-\lambda)t ) \le \rho(X),
\end{align}
which is clearly implied by quasi-star-shapedness. Hence, ``only-if" statement  holds true.
To show the ``if" statement, take arbitrary $X\in \X$ and $t\in \R$.
If $\rho(X)\le \rho(t)$, then we take $s\ge 0$ such that $\rho(X+s)=\rho(t)$. Such $s$ exists since $s\mapsto \rho(X+s)$ is continuous and $X+s\ge t$ for $s$ large enough.
Using  monotonicity of $\rho$ and \eqref{eq:oct-2}, we have
$$ 
 \rho(\lambda  X +(1-\lambda)t )  \le \rho(\lambda (X+s)+(1-\lambda)t ) \le \rho(t) =\max\{\rho(X),\rho(t)\}
$$ for each $\lambda\in [0,1]$. 
If $\rho(X)> \rho(t)$, then we take $s\ge 0$ such that $\rho(X )=\rho(t+s)$. Such $s$ exists since $s\mapsto \rho(t+s)$ is continuous and $t+s\ge X$ for $s$ large enough.
Using  monotonicity of $\rho$ and \eqref{eq:oct-2}, we have
$$ 
 \rho(\lambda  X +(1-\lambda)t )  \le \rho(\lambda X +(1-\lambda)(t+s) ) \le \rho(X)=\max\{\rho(X),\rho(t)\} 
$$ for each $\lambda\in [0,1]$.  Hence, quasi-star-shapedness holds. 
\end{proof}
\begin{remark}
 Proposition \ref{prop:oct-1} requires $L^\infty$-continuity, which is a weak property satisfied by essentially  all risk measures in the literature.
The result in Proposition \ref{prop:oct-1} holds true with the same proof if $L^\infty$-continuity is replaced by the property of \emph{solvability} in decision theory: 
For each $X\in \X$, there exists $t\in \R$ such that $\rho(X)=\rho(t)$. 
 \end{remark}

Proposition \ref{prop:oct-1} gives the following decision-theoretic interpretation of 
 quasi-star-shapedness.  
Suppose that the preference $\succeq$ of an agent satisfies \eqref{eq:oct-1}. If a random loss $X$ is seen as equally favourable as a constant loss $t$,
then $\lambda X+(1-\lambda)t$ is { weakly preferred to} $X$. That is, a combination of random $X$ and  constant $t$ reduces the riskiness of $X$.  
In contrast, quasi-convexity requires the above relation to hold for random $Y$ in {place} of constant $t$. Indeed, in the setting   of \cite{AA63} where $X$ and $Y$ represent acts with uncertainty (thus, they are not necessarily $\R$-valued),  the property,  for $X,Y\in \X$  and  $\lambda\in [0,1]$,  
\begin{align}\label{eq:oct-3}
X\simeq Y ~\Longrightarrow~  \lambda X+(1-\lambda)Y  \succeq X , 
\end{align}
is   the uncertainty aversion axiom of \cite{MMR06}, and it corresponds to quasi-convexity of the risk measure $\rho$ in our setting. 
It is clear that \eqref{eq:oct-1} is weaker than \eqref{eq:oct-3} as the riskiness of $X$ is only reduced when combined with an equally favourable constant loss, instead of an arbitrary   equally favourable loss $Y$. 

The difference between quasi-star-shapedness and  quasi-convexity, or between \eqref{eq:oct-1} and \eqref{eq:oct-3}, can also be explained via considerations for the dependence between pooled risks. 
For a law-invariant $\rho$ and two losses $X$ and $Y$ with fixed distributions, the dependence structure of $X$ and $Y$ affects   $\rho(\lambda X+ (1-\lambda )Y)$ but not $\rho(X)$ or $\rho(Y)$, and hence quasi-convexity  imposes  inequalities over all   dependence structures. Such an issue does not appear for $\lambda X+(1-\lambda)t$  as dependence is irrelevant between a random variable $X$ and a constant  $t$. 
Hence, relaxing quasi-convexity to quasi-star-shapedness gives rise to more flexibility on preferences over dependence. {In particular, under quasi-convexity, comonotonicity is the worst-case dependence in risk aggregation; see Lemmas \ref{prop:SSD} and \ref{lem:SSD}. This is not the case for quasi-star-shapedness, since $\VaR_t$ for $t\in (0,1)$ is quasi-star-shaped but it does not take comonotonicity as the worst-case dependence.
}
 
 Next, we discuss the issue of normalization. 
 The risk measures in Examples \ref{ex:put}   and \ref{exm:5} 
are  not necessarily normalized. In general,  cash-subadditive risk measures may not have the range of the entire real line.
Hence, normalization may also need to be weakened in our setting of cash subadditive risk measures, which we define as follows.
\begin{itemize}
\item[] \emph{Quasi-normalization}: $\rho(t)=t$ for all $t\in D_\rho$, where $D_\rho=\{\rho(X)\mid X\in \X\}$ is the range of $\rho$.
\end{itemize}
The risk measure $X\mapsto  \E[\min \{X,d\} ]$ in Example \ref{ex:put}  satisfies  quasi-normalization with range $(-\infty,d]$,
and $\Lambda\VaR$  in Example \ref{exm:5} satisfies quasi-normalization with range $(-\infty,z]$ where $z=\inf \{x\in \R:\Lambda(x)=0\}$ with the convention $\inf \emptyset =\infty$.

\subsection{A new representation of Lambda VaR}
The next result gives quasi-star-shapedness of $\Lambda\VaR$, complementing the fact  observed by \cite{CCMTW21} that 
 $\VaR$ is star-shaped.
   We also obtain, as a by-product, an alternative representation of $\Lambda\VaR$.
   In what follows, set $\VaR_0(X) =-\infty$ for any $X\in \X$, which follows from plugging $t=0$ in \eqref{eq:vardef}.

\begin{theorem}\label{prop:lambdarep}
Let $\Lambda: \R\to[0,1]$ be a decreasing function that is not constantly $0$.
The risk measure $\Lambda\VaR$ in \eqref{eq:lambdadef} has the representation
\begin{equation}
\label{eq:lambdarep}
\Lambda\VaR(X) = \inf_{x\in \R} \left\{\VaR_{\Lambda(x)} (X) \vee x\right\} 
= \sup_{x\in \R} \left\{\VaR_{\Lambda(x)} (X) \wedge x\right\} ,~~~~X\in \X,
\end{equation}
and moreover, $\Lambda\VaR$ is quasi-star-shaped.
\end{theorem}
\begin{proof} 
Note that for $X\in \X$, $x\in \R$ and $t\in [0,1]$, $P(X\le x) \ge t $ if and only if $\VaR_t(X)\le x$.
Moreover, since $\Lambda$ is decreasing,  the set $ \{x\in \R: \VaR_{\Lambda(x)}(X) \le  x\}$ is an interval with right end-point $\infty$. 
By definition, for $X\in \X$,
\begin{align*}
\Lambda\VaR(X) &= \inf \{x\in \R: P(X\le x) \ge \Lambda(x)\}
\\&  = \inf \{x\in \R: \VaR_{\Lambda(x)}(X) \le  x\} 
\\&  = \inf \{\VaR_{\Lambda(x)}(X) \vee  x: \VaR_{\Lambda(x)}(X) \le  x\} 
  \ge \inf_{x\in \R} \left\{ \VaR_{\Lambda(x)}(X) \vee  x \right\}.
\end{align*}  
 On the other hand, 
 \begin{align*}
\Lambda\VaR(X)  &  = \inf \{x\in \R: \VaR_{\Lambda(x)}(X) \le  x\} 
\\&  = \sup \{x\in \R: \VaR_{\Lambda(x)}(X)  >  x\} 
\\&  = \sup \{\VaR_{\Lambda(x)}(X) \wedge  x: \VaR_{\Lambda(x)}(X)>  x\} 
  \le \sup_{x\in \R} \left\{ \VaR_{\Lambda(x)}(X) \wedge  x \right\}.
\end{align*} 
Since   $\VaR_{\Lambda(x)}(X) \wedge  x  \le \VaR_{\Lambda(y)}(X) \vee  y $ for any $x,y\in \R$, we have 
$$
\Lambda\VaR(X)   \le \sup_{x\in \R} \left\{ \VaR_{\Lambda(x)}(X) \wedge  x \right\} 
\le \inf_{x\in \R} \left\{ \VaR_{\Lambda(x)}(X) \vee  x \right\} \le \Lambda\VaR(X) ,
$$
thus showing \eqref{eq:lambdarep}. Next, verify that the mapping $ X\mapsto \VaR_\alpha(  X)\vee x$ is quasi-star-shaped for all $\alpha\in [0,1]$ and $x\in\R$.
Note that if $\alpha=0$ then  it is trivial.
If $\alpha>0$, then $\VaR_\alpha(  X)\vee x =\VaR_\alpha(  X\vee x)$; {see Lemma A.27 of  \cite{FS16}}.
For all $X\in \X$, $t\in \R$ and $\lambda \in [0,1]$,
\begin{align}
  \VaR_\alpha(x\vee (\lambda X +(1-\lambda)t )) &\le  
\VaR_\alpha(\lambda ( x\vee  X)  +(1-\lambda)  ( x\vee  t) )   \notag
\\&=\lambda \VaR_\alpha(  x\vee  X)  +(1-\lambda)  \VaR_\alpha ( x\vee  t)   \notag
\\ &\le   \max\{\VaR_\alpha(x\vee X),\VaR_\alpha(x\vee t)\}.  \label{eq:lambdastar}
\end{align} 
Finally, we need to use  Lemma \ref{lem:weaknorm}  below, which states that the infimum of quasi-normalized, quasi-star-shaped and cash subadditive risk measures is quasi-star-shaped. 
Since $ X\mapsto \VaR_{\Lambda(x)}(X)\vee x$ is quasi-normalized, quasi-star-shaped and cash subadditive for all $x\in \R$,    $\Lambda\VaR$ is quasi-star-shaped by Lemma \ref{lem:weaknorm}. 
\end{proof}

\begin{remark}\label{rem:notstar}
We note that  $\Lambda\VaR$ is generally  not   star-shaped. 
For instance, take $\Lambda: x\mapsto \id_{\{x\le 1\}}$. 
For this choice, we have $\Lambda \VaR(x)=x \wedge 1$ for $x\in \R$.
It follows that $\Lambda \VaR(1)=1>1/2=\Lambda \VaR(2)/2+\Lambda \VaR(0)/2$, and hence $\Lambda \VaR$ is not star-shaped. 
Indeed, any $\Lambda$ with $\inf\{x\in \R: \Lambda(x)=0\} =1$ suffices for this example.
Note that each $X\mapsto \VaR_{\alpha}(X)\vee x$ in the representation \eqref{eq:lambdarep} is star-shaped (including $\alpha=0$); see \eqref{eq:lambdastar}. 
Therefore, the infimum of quasi-normalized, star-shaped and cash-subadditive risk measures is not necessarily star-shaped, in sharp contrast to the corresponding result on quasi-star-shaped risk measures in Lemma \ref{lem:weaknorm}. This example shows that quasi-star-shapedness is   more natural than, and genuinely different from, star-shapedness in the context of cash-subadditive risk measures.\footnote{On the other hand, for an increasing $\Lambda$,  $\Lambda\VaR(X)$  is in general not quasi-star-shaped.  A counter-example can be built using a Bernoulli random variable; see Example \ref{exm:non-qqs} in Appendix \ref{app:A5}.}

\end{remark}

Theorem \ref{prop:lambdarep} can be applied to solve portfolio optimization problems with $\Lambda\VaR$ constraints. Let $\Lambda:\R\to[0,1]$ be a   decreasing function which is not constantly $0$.
In a portfolio optimization problem, one often maximizes an objective, e.g., an expected utility or an expected return, under the constraint that 
a risk measure does not exceed a certain level $z$ (and often together with a budget constraint). 
 For $X\in\X$, by Theorem \ref{prop:lambdarep}, we have
$$\Lambda\VaR(X)\le z\iff\inf_{x\in\R}\{\VaR_{\Lambda(x)}(X) \vee x\}\le z\iff\inf_{x\le z}\VaR_{\Lambda(x)}(X)\le z\iff \VaR_{\Lambda(z)}(X)\le z.$$
Therefore, optimization under a $\Lambda\VaR$ constraint below a constant level $z$ is equivalent to that under a $\VaR_{\Lambda(z)}$ constraint below the same level $z$, which has been well studied in the risk management literature; see e.g., \cite{BS01} and \cite{BST06}.

 \subsection{A few useful technical results}
The next lemma shows that quasi-normalization and quasi-star-shapedness are preserved under a minimum operation, a fact used in the proof of Theorem \ref{prop:lambdarep}.
\begin{lemma}\label{lem:weaknorm}
The infimum of quasi-normalized, quasi-star-shaped, and cash-subadditive risk measures (assuming it is real-valued) is again quasi-normalized, quasi-star-shaped and cash-subadditive.
\end{lemma}
\begin{proof}
Let $\mathcal{C}$ be a class of quasi-normalized, quasi-star-shaped and cash-subadditive risk measures, and denote by $\rho=\inf_{\psi\in\mathcal{C}} \psi$.
It is obvious that $\rho$ is cash subadditive and monotone. It remains to show that $\rho$ is quasi-normalized and quasi-star-shaped.  
Denote by $d=\inf D_{\rho}$,  $u=\sup D_{\rho}$,
  $d_\psi = \inf D_{\psi}$ and $u_\psi=\sup D_{\psi}$ for $\psi \in \mathcal C$.   
For any $X\in \X$ and $\psi\in\mathcal C$,  if $u < d_{\psi}$, then $\rho (X) \le u < \psi(X)$. Hence, we can write  
$$\rho(X) = \inf_{\psi\in \mathcal C'} \psi (X)  \mbox{~where~}\mathcal C'=\{\psi \in \mathcal C \mid u\ge d_\psi\}.$$  
Note that $d\le d_\psi \le u\le u_\psi$ for each $\psi\in \mathcal C'$. 
{Moreover, by monotonicity and quasi-normalization of $\psi$,
for any $\psi \in \mathcal{C}  \cup\{\rho\}$, we have 
\begin{align}
& t \leq u_\psi~ \Longrightarrow ~\psi(t)=t \vee d_\psi,\label{eq:r1-2}\\
&t \geq d_\psi~ \Longrightarrow ~\psi(t)=t \wedge u_\psi. \label{eq:r1-1}
\end{align} 
 
We first show that $\rho$ is quasi-normalized.  Take a constant {$t\in (d,u)$}. 
Since $t< u\le u_\psi$, by \eqref{eq:r1-2}, we have $\psi(t)\ge t$ for all $\psi\in \mathcal C'$. Hence,
 $ 
\rho(t) = \inf_{\psi\in \mathcal C'} \psi (t)  \ge  t.
 $
Moreover, since $t>d$ and $d=\inf_{\psi\in \mathcal C'} d_\psi$, there exists $\psi\in \mathcal C'$ such that $d_\psi <t$. 
By \eqref{eq:r1-1},  we get $\psi(t) \le t$.}
Hence,
 $ 
\rho(t) = \inf_{\psi\in \mathcal C'} \psi (t)  \le  t.
 $
Thus, we obtain $\rho(t)=t$ for {$t\in (d,u)$}.  
It remains to verify $\rho(d)=d$ {(resp.~$\rho(u)=u$)} if $d>-\infty$ {(resp.~$u<\infty$)}.
This follows from the fact that a cash-subadditive risk measure is $L^\infty$-continuous. Therefore,  $\rho(t)=t$ for $t\in D_\rho$, and thus $\rho$ is quasi-normalized.
 
%

Next, we show that $\rho$ is quasi-star-shaped. 
 For $X\in \X$, $t\in\R$ and $\lambda \in [0,1]$, quasi-star-shapendess of $\psi\in \mathcal C'$ gives 
\begin{align}\label{eq:QCrho}
\rho(\lambda X+(1-\lambda) t)= \inf_{\psi\in \mathcal C'} \psi (\lambda X+(1-\lambda) t)\le \inf_{\psi\in \mathcal C'}\max\{ \psi (  X) ,\psi(t)\}.
\end{align}
If $t\ge u$, then $\rho(t)=u$ and  
$$\rho(\lambda X+(1-\lambda) t) \le u =\rho(t).$$
If $t<u$,
then   $\psi(t) =    t\vee d_{\psi}$  for  $\psi\in \mathcal C'$ and $\rho(t)=t \vee d$.
It follows that
\begin{align*}
\inf_{\psi\in \mathcal C'}\max\left\{ \psi (  X) ,\psi(t)\right\} &=  \inf_{\psi\in \mathcal C'}\max\left\{ \psi (  X) ,t, d_\psi \right\} \\&=  \inf_{\psi\in \mathcal C'}\max\left\{ \psi (  X) ,t \right\}  =  \max\left\{\inf_{\psi\in \mathcal C'} \psi (  X) ,t\right\} \le  \max\left\{\rho(  X) ,\rho(t)\right\} .
\end{align*}
Using \eqref{eq:QCrho} and combining both cases, we obtain $\rho(\lambda X+(1-\lambda)t)\le\max\{\rho(X),\rho(t)\}$ for all $\lambda\in[0,1]$, $X\in\X$ and $t\in\R$ and thus $\rho$ is quasi-star-shaped. 
\end{proof}

Finally, we show that in the classic setting of cash-additive risk measures, we do not need to distinguish between each of normalization,
star-shapedness and covexity and their quasi-versions. 
This result further illustrates that quasi-star-shapedness is a natural property to consider for cash-subadditive risk measures.  

\begin{proposition}
For  cash-additive risk measures,
\begin{enumerate}[(i)]
\item normalization is equivalent to quasi-normalization;
\item star-shapedness is equivalent to quasi-star-shapedness;
\item  convexity is equivalent to quasi-convexity. 
\end{enumerate}
In contrast, for cash-subadditive risk measures, none of the above equivalence  holds true.
\end{proposition}
\begin{proof}
The  statements on normalization are straightforward. 
Those on convexity are well known and can be checked with acceptance sets; see e.g., Proposition 2.1 and Example 2.2 of \cite{CMMM11}.
We only show the   statements on star-shapedness. 
\begin{enumerate}[(a)]
\item For cash-subadditive risk measures, the fact that these  star-shapedness and quasi-star-shapedness are not necessarily equivalent is illustrated in Remark \ref{rem:notstar}. 
\item  Suppose that   a cash-additive risk measure $\rho$ is star-shaped. Cash additivity and star-shapedness yield that,  for all $X\in \X$, $t\in \R$  and $\lambda\in[0, 1]$, 
$$\rho(\lambda X + (1-\lambda)t )=\rho(\lambda X)     +(1-\lambda)t  \le\lambda\rho(X) +(1-\lambda)\rho(t)\le \max\{\rho(X),\rho(t)\},$$ which implies that $\rho$ is quasi-star-shaped.
\item Suppose that  a cash-additive risk measure $\rho$  is quasi-star-shaped.
Let $\tilde \rho=\rho-\rho(0)$, and hence $\tilde  \rho$ is normalized. 
The acceptance set of $\tilde \rho$ is given by 
$$
\mathcal A_{\tilde \rho} = \{X\in \X: \tilde\rho(X)\le 0\}.
$$
Note that $0\in \mathcal A_{\tilde \rho}$ and $\tilde \rho$ is quasi-star-shaped. 
Therefore, for any $X\in  \mathcal A_{\tilde \rho}$ and $\lambda \in [0,1]$, we have
$\tilde \rho(\lambda X) \le \max \{\tilde \rho(X),\rho(0)\}\le 0$.
Hence, $\lambda X \in \mathcal A_{\tilde \rho}$,
and thus the set $\mathcal A_{\tilde \rho}$ is star-shaped. By Proposition 2  of \cite{CCMTW21}, we know that $\tilde \rho$ is star-shaped. In turn, this implies that $\rho$ is star-shaped. \qedhere
\end{enumerate}
\end{proof}


{\subsection{Quasi-star-shapedness of other cash-subadditive risk measures}
In addition to $\Lambda$VaR, we discuss quasi-star-shapedness of other risk measures in the examples in Section \ref{sec:pre}.  
 First, we consider the expected loss $\rho:X\mapsto \E [f(X)]$ in  Example \ref{ex:put} {and obtain a characterization result stronger than the second statement of Lemma 5.2 in \cite{CMMM11}}.  
Note that  $1$-Lipschitz continuity of $f$ is equivalent to cash subadditivity of $\rho$, but for this result we only need continuity.
\begin{proposition} \label{prop:ex22}
Suppose that  $f$ is  continuous and  increasing function on $\R$. 
The   expected loss
 $\rho:X\mapsto \E[f(X)]$  on $\X$ is quasi-star-shaped if and only if $f$ is  convex.
\end{proposition}
\begin{proof} 
The ``if"  statement is straightforward, as convexity is stronger than quasi-star-shapedness.
 
Below, we will show the ``only if" statement. 
For convexity, it suffices to show $f((a+b)/2)\le (f(a)+f(b))/2$ for all $a<b$ since $f$ is continuous.
We prove this in a few steps. First, we show that if $f$ has a positive derivative, then
the conclusion holds true. In the second step, we show that $f$ is either strictly increasing or 
first a constant then strictly increasing. 
In the third step, we show that the conditions in Step 2 is sufficient to use the conclusion in Step 1.

{\bf Step 1.}  Define the set
\begin{align*}
\mathcal D=\{(a,b,t)\in\R^3:~a<t<b,~f(a)+f(b)=2f(t),~f~\text{is differentiable at }t~{\rm and}~f'(t)>0\}.
\end{align*}
For $(a,b,t)\in\mathcal D$, take a random variable $X$ with distribution specified by $\p(X=a)=\p(X=b)=1/2$. Quasi-star-shapedness implies for all $\lambda\in(0,1)$,
\begin{align*}
\frac{1}{2}( f(\lambda a+(1-\lambda)t)+f(\lambda b+(1-\lambda)t) ) =\E[f(\lambda X+(1-\lambda)t)]\le \max\{\E[f(X)],f(t)\}=f(t).
\end{align*}
Hence, we have
\begin{align*}
\frac{1}{t-a}\cdot\frac{f(\lambda b+(1-\lambda)t)-f(t)}{\lambda(b-t)}\le \frac{1}{b-t}\cdot\frac{f(t)-f(\lambda a+(1-\lambda)t)}{\lambda(t-a)}.
\end{align*}
Letting $\lambda\to0$ yields 
\begin{align*}
\frac{f'(t)}{t-a}\le \frac{f'(t)}{b-t}.
\end{align*}
It follows from $f'(t)>0$ that $a+b\le 2t$. By the monotonicity of $f$, we have 
\begin{align}\label{eq-cx}
\frac{1}{2}( f(a)+f(b)) =f(t)\ge f\left(\frac{a+b}{2}\right).
\end{align}
This completes the proof of Step 1.

{\bf Step 2.} Next, we  verify that for $a,b\in\R$ and $a<b$, $f(a)=f(b)$ implies $f(x)=f(a)$ for all $x<a$. We prove this by contradiction. Suppose that $f(a)=f(b)$ for some $a<b$ and $f$ is not a constant on $(-\infty,a]$. Since $f$ is increasing, there exists $x_0\in(-\infty,a)$ such that $f(x_0+\epsilon)>f(x_0)$ for any $\epsilon>0$. {Such an $x_0$ exists because we can always take $\tilde x\in(-\infty,a)$ with $f(\tilde x)<f(a)$ and choose $x_0=\sup\{x\in(-\infty,a):f(x)=f(\tilde x)\}<a$ due to the continuity of $f$. As a result, it is clear that $f(x_0)<f(a)$.} Take $X$ with $\p(X=x_0)=\p(X=b)=1/2$. The continuity of $f$ guarantees $2f(t)=f(x_0)+f(b)$ for some $t\in(x_0,a)$. Using quasi-star-shapedness, we have for $\lambda\in(0,1)$,
\begin{align*}
\frac{1}{2} ( f(\lambda x_0+(1-\lambda)t)+f(\lambda b+(1-\lambda)t) ) =\E[f(\lambda X+(1-\lambda)t)]\le \max\{\E[f(X)],f(t)\}=f(t).
\end{align*}
Setting $\lambda=(a-t)/(b-t)\in(0,1)$ yields
\begin{align*}
\frac{1}{2} ( f(x_0+(1-\lambda)(t-x_0))+f(a) ) \le f(t)=\frac{1}{2}(f(x_0)+f(b))=\frac{1}{2}( f(x_0)+f(a)) .
\end{align*}
Note that $f(x_0+(1-\lambda)(t-x_0))>f(x_0)$ because $t>x_0$, and this yields a contradiction. Hence, we have proved Step 2.

{\bf Step 3.} As shown in Step 2, we know that there are two possible cases: (i) $f$ is strictly increasing on $\R$; (ii) there exists $x_0\in\R$ such that $f(x)=f(x_0)$ for $x\le x_0$ and $f$ is strictly increasing on $[x_0,\infty)$. In the first case,  for $a,b\in\R$ and $a<b$, let $t=f^{-1}((f(a)+f(b))/2)$, where $f^{-1}$ is the inverse function of $f$. Since $f$ is strictly increasing and continuous, we have $t\in(a,b)$ and $2f(t)=f(a)+f(b)$. If $f$ is differentiable at $t$, then it follows from Step 1 that $f((a+b)/2)\le (f(a)+f(b))/2$. If $f$ is not differentiable at $t$, there exists a sequence $\{t_n\}_{n\in\N}\subseteq(a,b)$ such that $t_n\to t$ and $f$ is differentiable at $t_n$ for all $n\in\N$ because the monotonicity of $f$ implies that $f$ is differentiable almost everywhere. Define $b_n=f^{-1}(2f(t_n)-f(a))$. It is not difficult to verify that $b_n>t_n>a$ and $f(b_n)=2f(t_n)-f(a)$ for all $n\in\N$, and $b_n\to b$. Using Step 1, we have $f((a+b_n)/2)\le (f(a)+f(b_n))/2$ for all $n\in\N$, and the continuity yields $f((a+b)/2)\le (f(a)+f(b))/2$. In the second case, the above argument gives $f((a+b)/2)\le (f(a)+f(b))/2$ for $b>a\ge x_0$, and this is sufficient for the convexity of $f$.
\end{proof}

\begin{remark}
Consider 
the expected utility model 
with an increasing continuous  utility function $u$ and the preference relation  $\succeq$ on $\X$  given by $X\succeq Y \iff  \E[u(X)]\ge \E[u(Y)]$. Here $X$ and $Y$ are interpreted as financial surpluses.
Using Propositions \ref{prop:oct-1} and  \ref{prop:ex22},  $u$ is concave if and only if
$$
\mbox{for all $X\in \X$, $t\in \R$ and $\lambda \in [0,1]$:}~~~ X\simeq t 
 ~\Longrightarrow~ \lambda X + (1-\lambda) t \succeq  X.
$$
This gives a weaker sufficient condition for risk aversion in the expected utility model than the usual one where $t$ is replaced by a random variable $Y$; see \citet[Section 6]{PWW23} for a different relaxation of the usual condition.
\end{remark}
 
Next, we consider the certainty equivalent $\rho$ of $\alpha$-MEU in Example \ref{ex:alphaMEU}. For technical tractability, we assume that the discount factor is deterministic.
Such  $\rho$ is generally not quasi-convex or cash additive.

\begin{proposition}\label{prop:ex23} Let $\mathcal{Q}_1$ and $\mathcal{Q}_2$ be two nonempty, weak*-compact and convex sets of finitely additive probabilities. For $I\subseteq[0,1]$ and $\gamma>0$,
the risk measure  
\begin{equation}\label{eq:certainty}
\rho(X)=\sup_{r\in I}\left\{\frac{1}{\gamma}\log\left(\alpha \min_{Q_1\in\mathcal{Q}_1}\E_{Q_1}[\mathrm{e}^{\gamma r X}]+(1-\alpha)\max_{Q_2\in\mathcal{Q}_2}\E_{Q_2}[\mathrm{e}^{\gamma r X}]\right)\right\},~~X\in\X\end{equation}
is   quasi-star-shaped.
\end{proposition} 
\begin{proof} 
Take any $X\in\X$, $t\in\R$ and $\lambda\in[0,1]$.
Note that for any probability measure $Q$,  we have 
$$
\E_{Q}[\mathrm{e}^{\gamma r  \lambda X }] \le \left(\E_{Q}[\mathrm{e}^{\gamma r  X }]\right)^{\lambda}.
$$
This gives $\rho(\lambda X)\le \lambda \rho(X)$.
 We have 
  \begin{align*}& \rho(\lambda X+(1-\lambda) t) \\& =\sup_{r\in I}\left\{\frac{1}{\gamma}\log\left(\alpha\min_{Q_1\in\mathcal{Q}_1}\E_{Q_1}[\mathrm{e}^{\gamma r (\lambda X+(1-\lambda) t)}]+(1-\alpha)\max_{Q_2\in\mathcal{Q}_2}\E_{Q_2}[\mathrm{e}^{\gamma r (\lambda X+(1-\lambda) t)}]\right)\right\}\\
    & =\sup_{r\in I}\left\{\frac{1}{\gamma}\log\left(\alpha\min_{Q_1\in\mathcal{Q}_1}\E_{Q_1}[\mathrm{e}^{\gamma r \lambda X}]+(1-\alpha) 
  \max_{Q_2\in\mathcal{Q}_2}\E_{Q_2}[\mathrm{e}^{\gamma r \lambda X}]\right) + (1-\lambda) rt\right\}\\
  &\le  \rho(\lambda X)+\sup_{r\in I} \left\{(1-\lambda) rt\right\}
 \le \lambda \rho( X)+  (1-\lambda)   \rho(t).  \end{align*}
Therefore, quasi-star-shapedness (and star-shapedness) holds. 
\end{proof}
 We say a set $\mathcal A\subseteq\X$ is \emph{star-shaped} if  $\lambda X+(1-\lambda) t\in\mathcal A$ for all $\lambda\in[0,1]$, $X\in \mathcal A$ and $t\in \mathcal A \cap \R$. Next, we show that star-shapedness of the acceptance set $\mathcal A$ characterizes the quasi-star-shapedness of the risk measure $\rho_{\mathcal A, S}$ in Example \ref{exm:4}.
\begin{proposition}\label{prop:ex24}
    For a closed set $\mathcal A\subseteq\X$, the risk measure  $\rho_{\mathcal A,S}$ in \eqref{eq:elig} is  quasi-star-shaped if and only if $\mathcal A$ is star-shaped.   
\end{proposition}
 \begin{proof}
To show the ``only if"  statement, assume that $\rho_{\mathcal{A},S}$ is quasi-star-shaped. For all   $X\in \mathcal A$, $t\in \mathcal A \cap \R$, and  $\lambda \in[0,1]$,  $\rho_{\mathcal A, S}(\lambda X+(1-\lambda) t)\leq \max\{\rho_{\mathcal A, S}(X),\rho_{\mathcal A, S}(t)\}\leq 0$.  Hence,  $\lambda X+(1-\lambda) t\in \mathcal A$ which implies that   $\mathcal A$   is  star-shaped.    

Next, we show the ``if"  statement. 
 Suppose that $\mathcal A$ is star-shaped. For all  $X\in \X$, $t\in\R$, and $\lambda\in[0,1]$, we have
 $-\rho_{\mathcal A,S}(t)S_T/S_0+t\in\mathcal A\cap\R$ and $-\rho_{\mathcal A,S}(X)S_T/S_0+X\in \mathcal A$.
 The  star-shapedness of $\mathcal A$ implies that $\lambda X+(1-\lambda)t-\lambda\rho_{\mathcal A,S}(X)S_T/S_0 -(1-\lambda) \rho_{\mathcal A,S}(t)S_T/S_0 \in \mathcal A$. Thus,
 \begin{align*}\rho_{\mathcal A, S}(\lambda X+(1-\lambda)t) &= \inf \left\{m\in\R: \lambda X+(1-\lambda)t-\frac{m}{S_0}S_T\in \mathcal A \right \}\\
 &\leq \lambda\rho_{\mathcal A,S}(X)+(1-\lambda)\rho_{\mathcal A,S}(t)\leq\max\{\rho_{\mathcal A,S}(X),\rho_{\mathcal A,S}(t)\}.
 \qedhere
 \end{align*}

 \end{proof}

 Propositions \ref{prop:ex22}--\ref{prop:ex24} together illustrate that quasi-star-shapedness of 
 cash-subadditive risk measures appears under natural conditions, and it can hold in relevant cases where quasi-convexity does not hold (as in the cases of Propositions \ref{prop:ex23}--\ref{prop:ex24}). 
 
 }
 
\section{Representation results on cash-subadditive risk measures} \label{sec:3}





In this section, we present a representation result, Theorem \ref{thm:1}, of general cash-subadditive risk measures, which illustrates that a cash-subadditive risk measure is the lower envelope of a family of quasi-convex cash-subadditive risk measures.  


\begin{theorem}\label{thm:1} For a functional $\rho: \X\to \mathbb{R},$ the following statements are equivalent.\begin{enumerate}[(i)]
\item $\rho$ is a cash-subadditive risk measure.
\item There exists a set $\mathcal{C}$ of quasi-convex cash-subadditive risk measures  such that
\begin{equation}
\rho(X)=\min _{\psi\in\mathcal{C}} \psi(X), ~~\text { for all } X \in \X. \label{eq:repmain}
\end{equation} 

\end{enumerate}
\end{theorem}
In order to prove Theorem \ref{thm:1}, we need the following lemma, which will also be useful for a few other results.
\begin{lemma}\label{lem:represent}
If $\rho:\X\to\R$ is a risk measure, then $\rho(X)=\min_{Z\in\X}\rho_{Z}(X)$ for all $X\in\X$, where
$$\rho_{Z}(X)=\inf \{\rho(Z+m) \mid m \in \mathbb{R}, ~Z+m \geq X\}, ~~X,Z\in \X.$$
\end{lemma}
\begin{proof}
For all $X,Z\in\X$, by the definition of $\rho_{Z},$ we have
\begin{align*}
\rho_{Z}(X)=\rho(Z+\esssup (X-Z)).
\end{align*}
Since $\rho$ is monotone and  $Z+\esssup (X-Z)\ge X$,    we have $\rho_{Z}(X)=\rho(Z+\esssup (X-Z))\ge\rho(X)$.  Note that $\rho_X(X)=\rho(X+\esssup (X-X))=\rho(X)$. Thus,  we have  $\rho_Z(X)\ge\rho_X(X)$ and this gives 
$
\min _{Z \in \X} \rho_{Z}(X)=\rho_X(X)=\rho(X).
$
\end{proof}

\begin{proof}[Proof of Theorem \ref{thm:1}]
``(ii)  $\Rightarrow$ (i)" is obvious. We now prove ``(i) $\Rightarrow$ (ii)". Assume that $\rho$ is a cash-subadditive risk measure.
By Lemma \ref{lem:represent}, we have $\rho(X)=\min_{Z\in\X}\rho_Z(X)$ for all $X\in\X$, where
$$\rho_{Z}(X)=\inf \{\rho(Z+m) \mid m \in \mathbb{R}, Z+m \geq X\}=\rho(Z+\esssup (X-Z)), ~~X,Z\in\X.$$
It is clear that $\rho_Z$ is monotonic. We show that $\rho_{Z}$ is cash subadditive. Indeed, for all $m \ge 0$ and $X \in \X,$ we have 
\begin{align*}
\rho_{Z}(X+m) =\rho(Z+\esssup (X+m-Z))&=\rho(Z+\esssup (X-Z)+m) \\
& \leq \rho(Z+\esssup (X-Z))+m=\rho_{Z}(X)+m.
\end{align*}
Next, we show that $\rho_{Z}$ is quasi-convex. To this end, we need to show that, for all $\alpha \in \mathbb{R}$, 
$X_{1}, X_{2} \in \X \text { and } \lambda \in[0,1]$,
$$ 
\rho_{Z}\left(X_{i}\right) \leq \alpha,~ i=1,2 ~\Longrightarrow~ \rho_{Z}\left(\lambda X_{1}+(1-\lambda) X_{2}\right) \leq \alpha. 
$$
Assume that $\rho_{Z}\left(X_{i}\right) \leq \alpha$ for $i=1,2$. For all $\varepsilon>0$ and $i=1,2,$ there exists some $m_i \in \mathbb{R}$ such that $Z+m_i \geq X_{i}$ and $\rho\left(Z+m_i\right) \leq \rho_{Z}\left(X_{i}\right)+\varepsilon \leq \alpha+\varepsilon .$ Thus we have $$\lambda X_{1}+(1-\lambda) X_{2} \leq Z+\lambda m_1+(1-\lambda)m_2.$$
 It  then follows that
$$
\rho_{Z}\left(\lambda X_{1}+(1-\lambda) X_{2}\right) \leq \rho(Z+\lambda m_1+(1-\lambda)m_2) \leq \rho(Z+\max\{m_1,m_2\})\leq  \alpha+\varepsilon.
$$ 
The arbitrariness of $\varepsilon$ implies that $\rho_{Z}\left(\lambda X_{1}+(1-\lambda) X_{2}\right) \leq \alpha .$ Therefore, $\rho_{Z}$ is quasi-convex.
Finally, $\left\{  \rho_{Z}\mid Z \in \X \right\}$ is a desired family of quasi-convex cash-subadditive risk measures.
\end{proof}

The representation in Theorem \ref{thm:1} can be interpreted as that any cash-subadditive risk measure can be seen as a best-case representative from a collection of quasi-convex ones (which may be obtained through market competition, i.e., taking the cheapest price when risk measures are interpreted as price mechanisms), and this is similar to the situation in \cite{CCMTW21} in the context of monetary risk measures.
As far as we are aware of, Theorem \ref{thm:1} is the first characterization  result of cash-subadditive risk measures that are not necessarily quasi-convex.   A connection between Theorem \ref{thm:1}  and some results of \cite{JXZ21} on monetary risk measures 
are discussed in Appendix \ref{App:1}.
We also note that, by straightforward argument, an equivalent statement to Theorem \ref{thm:1} (ii) is 
\begin{equation}
\label{eq:general} \rho(X)=\min \{\psi(X) \mid \psi \text{ is a quasi-convex cash-subadditive risk measure},~\psi \geq \rho\},~~X\in\X.\end{equation}
Note that \eqref{eq:general} gives the largest set $\mathcal C $ of quasi-convex cash-subadditive risk measures for which the representation in  Theorem \ref{thm:1} holds.

 \begin{remark} Using the same argument as for Theorem \ref{thm:1}, a similar result  holds for risk measures without cash subadditivity; that is, a functional $\rho:\X\to\R$ is a risk measure if and only if
\begin{equation*}
\rho(X)=\min \{\psi(X) \mid \psi \text{ is a quasi-convex risk measure},~\psi \geq \rho\},~~ X\in\X.
\end{equation*}
\end{remark}

\begin{example}[$\Lambda\VaR$]
For a decreasing function $\Lambda:\R\to[0,1]$ that is not constantly $0$, by Theorem \ref{prop:lambdarep}, the $\Lambda\VaR$ in \eqref{eq:lambdadef} admits the representation
$\Lambda\VaR(X)=\inf_{x\in\R}\{\VaR_{\Lambda(x)}(X)\vee x\}$, $X\in\X$.
Since $\VaR$ commutes with continuous increasing transforms, we have
\begin{equation}
\label{eq:exlambda}
\Lambda\VaR(X)=\inf_{x\in\R}\VaR_{\Lambda(x)}(X\vee x).
\end{equation}
Let $\mathcal C_t$ be the set of coherent risk measures dominating $\VaR_t$ for $t\in (0,1)$. 
By  Theorem 6.8 of \cite{D02},  $\VaR_t$ has the representation
\begin{equation}
\label{eq:exlambda2}
\VaR_t(X) = \min_{\rho \in \mathcal C_t} \rho(X),~~~~X\in \X. \end{equation}
For $x\in\R$, denote by $\tau_x:X\mapsto \tau(X\vee x)$ for   $\tau\in \mathcal C_{\Lambda(x)}$ and by $\mathcal C_{\Lambda,x} = \{\tau_x: \tau \in \mathcal C_{\Lambda(x)}\}$. 
Using \eqref{eq:exlambda} and \eqref{eq:exlambda2}, we get the representation \eqref{eq:repmain} for $\Lambda \VaR$
as
\begin{equation}
\label{eq:exlambda3}
\Lambda\VaR (X)    = \min \left\{ \rho(X)~\middle |~ \rho\in \bigcup_{x\in\R} \mathcal C_{\Lambda,x}\right\},~~~~X\in \X.
\end{equation}
We check that $\tau_x\in\mathcal C_{\Lambda,x}$  is cash subadditive and quasi-convex for any $x\in\R$. Indeed,   $\tau_x$ is convex since it is the composition of a convex risk measure $\tau$ and a convex transform $y\mapsto y\vee x$. 
To see that it is cash subadditive, it suffices to note that $\tau_x(X+c) \le\tau(X\vee x +c)  = \tau_x ( X)+c$ for $c\ge 0$.

A special case of $\Lambda\VaR$ is the  two-level $\Lambda\VaR$  in Example 7 of \cite{BP21},  which is the simplest form of $\Lambda\VaR$  different from $\VaR$; we give a more explicit formula for this case. Fix $0<\alpha<\beta<1$ and $z\in \R$.  Define $\Lambda': x\mapsto \beta \id_{\{x\le z\}} + \alpha \id_{\{x>z\}}$. The corresponding risk measure is given by   $
\Lambda'\VaR (X) = \min\{\VaR_\beta(X), \VaR_\alpha(X\vee z)\}$, $X\in \X.
$  
Write $\mathcal C_{t,x} = \{\tau_x: \tau \in \mathcal C_t\}$ for $x\in\R$ and $t\in(0,1)$. 
By \eqref{eq:exlambda3},
$$
\Lambda'\VaR (X)    
= \min \{ \rho(X)\mid \rho\in \mathcal C_{\beta}\cup \mathcal C_{\alpha,z}\},~~~X\in \X.
$$
\end{example}
{  The representation  \eqref{eq:exlambda3} is  parallel to the property that $\VaR_t$ can be represented as the lower envelope the coherent risk measures dominating $\VaR_t$ in \eqref{eq:exlambda2}.  Since $\Lambda \VaR$ has a similar interpretation to $\VaR$ via assessing risks with loss probability, 
and  cash subadditivity and quasi-convexity generalize cash additivity and convexity, the representation \eqref{eq:exlambda3} arises quite naturally.}



Next, we look at a more explicit representation of cash-subadditive risk measures. 
An existing result  of \cite{CMMM11} states that a quasi-convex cash-subadditive risk measure can be represented by the supremum of a family of functions  $(t,Q)\mapsto R (t,Q)$ that are upper semi-continuous, quasi-concave, increasing and $1$-Lipschitz in its first argument $t$.  Combining Theorem \ref{thm:1} and Theorem 3.1 of \cite{CMMM11}, 
we   obtain a   representation of a general cash-subadditive risk measure based on the above functions $R$. 



\begin{proposition}\label{prop:class}
A functional $\rho:\X\to\R$ is a cash-subadditive risk measure if and only if there exists a set $\mathcal{R}$ of upper semi-continuous,  quasi-concave,  increasing and $1$-Lipschitz in the first argument functions $R: \mathbb{R} \times \mM_f \rightarrow\R$ such that 
\begin{equation*}
\rho(X) = \min_{R\in \mathcal R} \max_{Q\in\mM_f} R\left(\mathbb{E}_{Q}[X], Q\right),~~\text{for all }X\in \X.
\end{equation*}
\end{proposition}

Proposition \ref{prop:class} has a similar form to the minimax representation of star-shaped risk measures in Proposition 5 of \cite{CCMTW21}. 
We note that the set $\mathcal R$ is not unique in this representation, 
 different from $R$ in \cite{CMMM11}, which is unique. Instead, the largest choice of $\mathcal R$ is unique, which is the set of all $R$ satisfying the conditions in Proposition \ref{prop:class} such that 
$ \max_{Q\in\mM_f} R\left(\mathbb{E}_{Q}[X], Q\right)\ge \rho(X) $ for all $X$.



\section{Cash-subadditive risk measures with further properties}\label{sec:extend}


\subsection{Normalized and quasi-star-shaped cash-subadditive risk measures} \label{sec:PH}
In this section, {we give a representation   of cash-subadditive risk measures that are normalized and quasi-star-shaped in  Theorem \ref{thm:PH}.  Some other relevant technical results are also obtained.} Before showing Theorem \ref{thm:PH}, we need the representation result below of quasi-normalized, quasi-star-shaped and cash-subadditive risk measures, which is in similar sense with Lemma \ref{lem:represent} but based on a more sophisticated construction with techniques different from the literature. 
In what follows, the convention is $\sup\emptyset = -\infty$ so that  all quantities are  well defined.
\begin{proposition}\label{prop:rep_star}
Let $\rho:\X\to\R$ be a quasi-normalized, quasi-star-shaped and cash-subadditive risk measure. For $Z\in\X$ and $t\in\R$, define $$m_Z(t)=\sup\{m\in\R\mid\rho(Z+m)= t\}~~\text{and}~~\mathcal{A}^t_Z=\bigcup_{\lambda\in[0,1]}\{X\in\X\mid X\le \lambda(Z+m_Z(t))+(1-\lambda)t\}.$$ We have  $\rho(X)=\min_{Z\in\X}\tilde{\rho}_{Z}(X) $ for $X\in\X$, where
 $ \tilde{\rho}_{Z}(X)=\inf\{t\in\R\mid X\in\mathcal{A}^t_Z\}. $
\end{proposition}
\begin{proof} 

Since a cash-subadditive risk measure is $L^\infty$-continuous,  for each $Z\in \X$,
the range of the   function $m\mapsto \rho(Z+m)$ on $\R$ is an interval of $\R$. 
Moreover, recall the definition of $D_\rho=\{\rho(X)\mid X\in \X\}$, since $\rho$ is quasi-normalized, the function $m\mapsto \rho(m)$ on $\R$ takes all possible values in $D_\rho$, which is an interval on $\R$, and by monotonicity, so does $m\mapsto \rho(Z+m)$.
 Hence, 
$\rho(Z+m_Z(t))=t$ for all $t\in D_\rho$. 
For   $X,Z\in\X$, we can write
 \begin{align*}
\tilde{\rho}_Z(X)&=\inf\{t\in\R\mid X\in\mathcal{A}^t_Z\} \\&=\inf\{t\in\R\mid X\le \lambda(Z+m_Z(t))+(1-\lambda)t\text{ for some }\lambda\in[0,1]\}
\\& =\inf_{\lambda\in[0,1]}\inf\{t\in\R\mid X\le \lambda(Z+m_Z(t))+(1-\lambda)t\}
.
\end{align*} 

It is straightforward that $\tilde\rho_Z(X)\in D_\rho$.
For  $X,Z\in \X$ and $t\in D_\rho$, if $\tilde\rho_Z(X) < t$, then $X\le\lambda(Z+m_Z(t))+(1-\lambda)t$ for some  $\lambda\in[0,1]$. By monotonicity, quasi-normalization and quasi-star-shapedness of $\rho$, we have
$$\rho(X)\le\rho(\lambda(Z+m_Z(t))+(1-\lambda)t)\le\max\{\rho(Z+m_Z(t)),t\}=t.$$
Thus we have $\rho(X)\le\inf_{Z\in\X}\tilde{\rho}_Z(X)$. 
On the other hand,  
 \begin{align*}
 \tilde{\rho}_Z(X)&\le \inf\{t\in\R\mid X\le Z+m_Z(t)\}\\&=\inf\{t\in\R\mid m_Z(t)= \esssup(X-Z)\}  =\rho(Z+\esssup(X-Z)),
 \end{align*}  
 which is $\rho_Z(X)$ in Lemma \ref{lem:represent}. Using Lemma \ref{lem:represent}, we have \begin{equation}\label{eq:new-old-rep}\rho(X) = \min_{Z\in\X} {\rho}_Z(X)\ge \inf_{Z\in\X}\tilde{\rho}_Z(X) \ge \rho(X).\end{equation}
 Moreover, attainability of the infimum is guaranteed by $\rho(X)=\rho_X(X) \ge \tilde \rho_X(X)\ge \rho(X)$. 
Therefore, $\rho(X)=\min_{Z\in\X}\tilde\rho_Z(X)$ holds. 
\end{proof}

The representation in Proposition \ref{prop:rep_star} is closely linked to that in Lemma \ref{lem:represent} through \eqref{eq:new-old-rep}.  
\begin{remark}
Although arising from completely different considerations, the risk measure $\tilde \rho_Z$ in Proposition \ref{prop:rep_star}   has a similar form to an acceptability index of \cite{CM09}, {which has the form  
 $ 
\alpha(X)=\sup \left\{x \in \mathbb{R}_{+} \mid X \in \mathcal{A}_x\right\}
$  where  
$ 
(\mathcal{A}_x)_{x\in \R_+}
$ is a decreasing family of  subsets of $\X$.}  
For more recent results on acceptability indices, see e.g., \cite{R21}.
\end{remark}

The following representation result concerns cash-subadditive risk measures that are normalized and quasi-star-shaped. We show that
a normalized, quasi-star-shaped  and cash-subadditive risk measure can be represented by the lower envelope of a family of  ones that are normalized, quasi-convex, and cash subadditive.

\begin{theorem}\label{thm:PH}
For a functional $\rho: \X \rightarrow \mathbb{R},$ the following statements are equivalent.
\begin{enumerate}[(i)]
\item $\rho$ is a normalized, quasi-star-shaped and cash-subadditive risk measure.
\item There exists a family $\mathcal{C}$ of normalized, quasi-convex and cash-subadditive risk  measures  such that
\begin{equation}
\rho(X)=\min _{\psi\in\mathcal{C}} \psi(X), ~~\text { for all } X \in \X. \label{eq:repstar}
\end{equation} 
\end{enumerate}
\end{theorem}
\begin{proof}
``(ii)  $\Rightarrow$ (i)": Assume that there exists a family $\mathcal{C}$ of normalized, quasi-convex and cash-subadditive risk measures such that $\rho =\min_{\psi\in\mathcal{C}}\psi$. Monotonicity, normalization and cash subadditivity of $\rho$ are straightforward. 
Quasi-star-shapedness follows from Lemma \ref{lem:weaknorm}.

``(i) $\Rightarrow$ (ii)": Assume that $\rho$ is a normalized, quasi-star-shaped and cash-subadditive risk measure. Using Proposition \ref{prop:rep_star}, 
it suffices to show that  $\tilde \rho_Z(X)$ defined via  
$$\tilde{\rho}_Z(X)=\inf_{\lambda\in[0,1]}\inf\{t\in\R\mid X\le \lambda(Z+m_Z(t))+(1-\lambda)t\}$$
for each $Z\in \X$ is a normalized, quasi-convex and cash-subadditive risk measure.

We first verify that each $\tilde\rho_Z$ is normalized. For all $s\in\R$, by taking $\lambda=0$, we have $\tilde\rho_Z(s)\le\inf\{t\in\R\mid s\le t\}=s$. On the other hand, for all $t\in\R$ and $\lambda\in[0,1]$ such that $s\le\lambda(Z+m_Z(t))+(1-\lambda)t$, by normalization, monotonicity and quasi-star-shapedness of $\rho$, we have
$$s=\rho(s)\le\rho(\lambda(Z+m_Z(t))+(1-\lambda)t)\le\max\{\rho(Z+m_Z(t)),t\}= t.$$
Hence we obtain $\tilde\rho_Z(s)\ge s$, and further $\tilde\rho_Z(s)=s$.

Next, we show that each $\tilde\rho_Z$ is quasi-convex.
We first note that $\mathcal A_Z^t$ is a convex set for each $t\in \R$, which follows from the fact that $\mathcal A_Z^t$ is the set of all $X\in \X$ dominated by the segment 
$
\{\lambda(Z+m_Z(t))+(1-\lambda)t \mid \lambda \in [0,1] \}.
$
 Take $t \in\R$.
For $X_1,X_2$ satisfying $\tilde\rho_Z(X_1)\le \tilde\rho_Z(X_2)\le t$,
for any $s>t$, we have $X_1,X_2\in \mathcal A^s_Z$. Convexity of $\mathcal A^s_Z$ implies, for each $\lambda \in [0,1]$, 
$ \lambda X_1+(1-\lambda)X_2 \in  \mathcal A^s_Z$,
and it  further gives $\tilde \rho_Z(  \lambda X_1+(1-\lambda)X_2)\le s$.
Since $s>t$ is arbitrary, we have $\tilde \rho_Z(  \lambda X_1+(1-\lambda)X_2)\le t$.
This gives quasi-convexity of $\tilde \rho_Z$. 
%
%

Finally, we prove that $\tilde\rho_Z$ is cash subadditive for all $Z\in\X$.
{Note that continuity in Remark \ref{rem:continuity} implies 
$$
m_Z(t)=\sup\{m\in\R\mid\rho(Z+m)\le  t\}.
$$
 Since $\rho$ is cash subadditive, for all $Z\in\X$, $t\in\R$ and $c\ge 0$,
\begin{equation*}
\begin{aligned}
m_Z(t+c)&=\sup\{m+c\in\R\mid\rho(Z+m+c)\leq t+c\}\\
&\ge\sup\{m+c\in\R\mid\rho(Z+m)+c\leq t+c\}=m_Z(t)+c.
\end{aligned}
\end{equation*}}
For all $c\ge 0$ and $X\in\X$, we have
$$\begin{aligned}
\tilde\rho_Z(X+c)&=\inf_{\lambda\in[0,1]}\inf\{t\in\R\mid X+c\le \lambda(Z+m_Z(t))+(1-\lambda)t\}\\
&=\inf_{\lambda\in[0,1]}\inf\{t+c\in\R\mid X+c\le \lambda(Z+m_Z(t+c))+(1-\lambda)(t+c)\}\\
&=\inf_{\lambda\in[0,1]}\inf\{t\in\R\mid X\le \lambda(Z+m_Z(t+c)-(t+c))+t\}+c\\
&\le\inf_{\lambda\in[0,1]}\inf\{t\in\R\mid X\le \lambda(Z+m_Z(t)-t)+t\}+c=\tilde\rho_Z(X)+c.
\end{aligned}$$
In summary,  
$\{\tilde{\rho}_Z\mid Z\in \X\}$ is a desired family of normalized, quasi-convex and cash-subadditive risk measures.
\end{proof}

The proof of Theorem \ref{thm:PH} is based on a delicate construction of the dominating risk measures, different from those used for Theorem \ref{thm:1}.
Normalization in both (i) and (ii) of  Theorem \ref{thm:PH} is important and cannot be removed, but it can be replaced by quasi-normalization. 
The modified version of Theorem \ref{thm:PH} using quasi-normalization follows from combining Proposition \ref{prop:rep_star} and Lemma \ref{lem:weaknorm}.


Theorem \ref{thm:PH} can be seen as  a parallel result, although obtained via different techniques,  to  the representation result of \cite{CCMTW21},  which uses   star-shapedness, convexity, and cash additivity instead of  quasi-star-shapedness, quasi-convexity and cash subadditivity. 
It is clear that (ii) of  Theorem \ref{thm:PH} is equivalent to the following alternative formulation
\begin{align}
\rho(X)=\min \left\{\psi(X) ~\middle | \begin{array}{l} \psi \text{ is a normalized, quasi-convex and}\\
\text{cash-subadditive risk measure},~ \psi \geq \rho \end{array}\right\},~~X\in \X.\label{eq:repstar2}
\end{align}

\subsection{SSD-consistent cash-subadditive risk measures} \label{sec:SSD}

{Let $(\Omega, \mathcal{F}, P)$ be a nonatomic probability space} in this section. We present below the representation result of SSD-consistent cash-subadditive risk measures.  For this, we  define the \emph{Expected Shortfall} (ES)   at level  $t\in[0,1]$ as
\begin{align*}
\ES_{t}(X)=\frac{1}{1-t} \int_{t}^{1} \VaR_{\alpha}(X) \,\mathrm{d}\alpha,~t\in[0,1)~~\mbox{and}~~\ES_1(X)=\esssup(X),~~X\in \X.
\end{align*}
As a coherent alternative to VaR,
ES is the most important risk measure in current banking regulation; see \cite{WZ21} for its role in the Basel Accords and an axiomatization.
It is well known that the class of ES characterizes SSD via
\begin{align*}
X \succeq_{2} Y\iff \ES_t(X)\ge\ES_t(Y) \text{ for all }t\in[0,1].
\end{align*}

\cite{MW20} investigated SSD-consistent monetary risk measures and provided four equivalent conditions of SSD-consistency; see their Theorem 2.1. 
The result  can also be extended to {$L^\infty$-continuous} risk measures, which is shown in the following lemma. 
\begin{lemma}\label{prop:SSD}
Let $\rho$ be  an {$L^\infty$-continuous} risk measure  on $\X$. The  following are equivalent.
 \begin{enumerate}[(i)]
\item $\rho$ is SSD-consistent.
\item  $\rho(X) \geq \rho(Y)$ for all $X,Y\in\X$ with  $X \succeq_{2} Y$
 and $\mathbb{E}[X]=\mathbb{E}[Y]$.
\item $\rho(X) \geqslant \rho(Y)$ for all $X,Y\in\X$ with  $\mathbb{E}\left[(X-K)_{+}\right] \geqslant \mathbb{E}\left[(Y-K)_{+}\right]$ for all $K \in \mathbb{R}$. 
\item  $\rho(X) \geqslant \rho(Y)$ for all $X,Y\in\X$ with $Y=\mathbb{E}[X \mid Y]$.
\item   $\rho(X^{c}+Y^{c})\geqslant \rho(X+Y) $ for all $X,Y,X^c,Y^c\in\X$ such that $(X^{c}, Y^{c})$ is comonotonic, $X \stackrel{\mathrm{d}}{=} X^{c}$, and
$Y \stackrel{\mathrm{d}}{=} Y^{c}$.

\end{enumerate}  Moreover,    any of these properties imply that $\rho$ is law invariant.
\end{lemma}
\begin{proof} {Since $(\Omega, \mathcal{F}, P)$ is nonatomic,} the equivalence among (i)-(iv) is easy to verify from   classic properties of SSD  by the same logic of the proof of  Theorem 2.1  in  \cite{MW20}.    The equivalence between (i) and (v) for $L^\infty$-continuous functions  follows from Theorem 2 of \cite{WW20}. 
\end{proof}

The following lemma is needed in the proof of Theorem \ref{thm:SSD}, which was obtained by \cite{CMMM11} with the additional assumption of continuity from above. We include a self-contained proof of Lemma \ref{lem:SSD}.

\begin{lemma}\label{lem:SSD}
If  $\rho:\X\to\R$ is a quasi-convex cash-subadditive  risk measure, then $\rho$ is law invariant if and only if $\rho$ is SSD-consistent.
\end{lemma}
\begin{proof} 
It is obvious that SSD-consistency implies law invariance. We will only show the ``only if" statement. 
By Lemma \ref{prop:SSD}, it suffices to show that  $\rho(X)\le \rho(Y)$ for  $X\preceq_{\rm 2} Y$ satisfying $\E[X]=\E[Y]$.
By Proposition 3.6 of \cite{MW15}, there exists a sequence of $\mathbf Y^k=(Y_1^k,\dots,Y_{n_k}^k)$, $k\in \N$, such that each $Y_j^k\laweq Y$,  $n_k\to \infty$ as $k\to \infty$, and 
$$
\frac{1}{n_k} \sum_{j=1}^{n_k} Y^k_j \to X  \mbox{~~in $L^\infty$}.
$$
Note that a cash-subadditive risk measure is $L^\infty$-continuous. 
Quasi-convexity  and $L^\infty$-continuity lead to 
$$
 \rho(Y) \ge \rho\left (\frac{1}{n_k} \sum_{j=1}^{n_k} Y^k_j  \right) \to \rho(X),
$$
and thus $\rho$ is SSD-consistent.
\end{proof}

In the following theorem, we establish a representation for an SSD-consistent  cash-subadditive risk measure as the lower envelope of some family of law-invariant, quasi-convex and  cash-subadditive risk measures. 
 
\begin{theorem}\label{thm:SSD}
For a functional $\rho: \X \rightarrow \mathbb{R},$ the following statements are equivalent.
\begin{enumerate}[(i)]
\item $\rho$ is an SSD-consistent cash-subadditive risk measure.
\item There exists a family $\mathcal{C}$ of law-invariant, quasi-convex and cash-subadditive risk  measures  such that
$$
\rho(X)=\min _{\psi\in\mathcal{C}} \psi(X), ~~\text { for all } X \in \X.
$$ 
\end{enumerate}
\end{theorem} 

\begin{proof}
  ``(ii) $\Rightarrow$ (i)"  is implied by Lemma \ref{lem:SSD} (i) and the fact that cash subadditivity and SSD-consistency are preserved under the infimum operation. We will show ``(i) $\Rightarrow$ (ii)".

Suppose that $\rho$ is an  SSD-consistent cash-subadditive risk measure. For all $X\in\X$ and $Z\in \X$, define the risk measure
$$
\psi_{Z}(X)=\inf \{\rho(Z+m) \mid m \in \mathbb{R},~ Z+m\succeq_{2}  X\}.
$$
It is straightforward to check that $\rho(X)=\min_{Z\in\X}\psi_Z(X)$ and
$$\begin{aligned}
\psi_{Z}(X)&=\inf \{\rho(Z+m) \mid m \in \mathbb{R},~ \ES_t(Z)+m\ge  \ES_t(X),~\text{for all }t\in[0,1]\}\\
&=\rho\left(Z+\sup_{t\in[0,1]}(\ES_t(X)-\ES_t(Z))\right).
\end{aligned}$$
It is clear that $\psi_Z$ is monotone, cash subadditive and law invariant. We prove that $\psi_Z$ is quasi-convex with similar manner to the proof of Theorem \ref{thm:1}. Assume that $\psi_{Z}\left(X_{i}\right) \leq \alpha$ for $i=1,2$. For all $\varepsilon>0$ and $i=1,2,$ there exists some $m_i \in \mathbb{R}$ such that $Z+m_i \succeq_2 X_{i}$ and $\rho\left(Z+m_i\right) \leq \psi_{Z}\left(X_{i}\right)+\varepsilon \leq \alpha+\varepsilon $. { By Theorem 3.5 of \cite{R13}},  we have
$Z+\lambda m_1+(1-\lambda)m_2\succeq_2 \lambda X_{1}+(1-\lambda) X_{2}$  for all $\lambda\in[0,1]$.
It follows that
$$
\psi_{Z}\left(\lambda X_{1}+(1-\lambda) X_{2}\right) \leq \rho(Z+\lambda m_1+(1-\lambda)m_2) \leq\rho(Z+\max\{m_1,m_2\}) \leq \alpha+\varepsilon.
$$ 
The arbitrariness of $\varepsilon$ implies that $\psi_{Z}\left(\lambda X_{1}+(1-\lambda) X_{2}\right) \leq \alpha .$ Therefore, $\psi_{Z}$ is quasi-convex.
We conclude that $\left\{  \psi_{Z}\mid Z \in \X \right\}$ is a desired family of law-invariant, quasi-convex and cash-subadditive risk measures. 
\end{proof} 

Theorem \ref{thm:SSD} can be seen as a parallel result to Theorem 3.3 of \cite{MW20} which showed that 
any SSD-consistent monetary risk measure is the lower envelope of law-invariant and convex monetary risk measures. 
 Similarly to \eqref{eq:repstar2}, we can reformulate (ii) of Theorem \ref{thm:SSD} as 
\begin{align}
\rho(X)=\min \left\{\psi(X) ~\middle | \begin{array}{l} \psi \text{ is a law-invariant, quasi-convex and}\\
\text{cash-subadditive risk measure},~ \psi \geq \rho \end{array}\right\},~~X\in \X.\label{eq:equallaw}
\end{align}

A   representation result  in a similar spirit to Proposition \ref{prop:class}  
for SSD-consistent cash-subadditive risk measures follows directly from Theorem 5.1 of \cite{CMMM11} and Theorem \ref{thm:SSD}. 
\begin{proposition}
{ Let $\rho:\X\to\R$ be a functional that is continuous from above.} We have $\rho$ is an SSD-consistent cash-subadditive risk measure if and only if there exists a set $\mathcal{R}$ of upper semi-continuous, quasi-concave, increasing and $1$-Lipschitz in the first component functions $R: \mathbb{R} \times {\mM} \rightarrow\R$ such that 
$$ \rho(X) = \min_{R\in \mathcal R} \max_{Q\in{\mM}} R\left(\int_{0}^{1} \VaR_t(X) \VaR_t\left({\frac{\mathrm{d} Q}{\mathrm{d} P}}\right) \d t, ~Q\right),~~\text{for all }X\in \X.$$
\end{proposition}

\section{Conclusion}\label{sec:6}

We provide a systemic study of cash-subadditive
risk measures, which were traditionally studied together with convexity \citep{KR09} or quasi-convexity \citep{CMMM11}.
Different from the literature,  
our study focuses on cash-subadditive risk measures without quasi-convexity, which include many natural examples as discussed in the paper. 
As our major technical contributions, a general cash-subadditive risk measure is shown to be representable by the lower envelope of a family of quasi-convex cash-subadditive risk measures (Theorem \ref{thm:1}). The notions of quasi-star-shapedness and quasi-normalization were introduced as analogues of star-shapedness and normalization studied by \cite{CCMTW21}. It turns out that quasi-star-shapedness and quasi-normalization fit naturally in the setting of cash subadditivity, leading to a new representation result (Theorem \ref{thm:PH}). A representation result of SSD-consistent cash-subadditive risk measures was also obtained (Theorem \ref{thm:SSD}). 
We summarize some related results in the literature and compare them with our results in Table \ref{tab:1}. 

\begin{table}[htbp] 
\def\arraystretch{1.8}
  \begin{center} \small
    \begin{tabular}{l|l|l} 
       & a (...) risk measure  &  is an infimum of (...) risk measures  \\ \hline
    \cite{MW20} & CA, SSD-consistent    & CA,  convex, law-invariant  \\ \hline
    \cite{JXZ21} & CA  & CA, convex    \\ \hline
    \cite{CCMTW21} & CA, star-shaped, normalized & CA,   convex, normalized \\ \hline
    Theorem \ref{thm:SSD}   & CS, SSD-consistent &   CS, quasi-convex, law-invariant \\ \hline 
    Theorem \ref{thm:1} & CS &   CS, quasi-convex  \\ \hline
    Theorem \ref{thm:PH}  & CS, quasi-star-shaped, normalized &  CS, quasi-convex, normalized\\ \hline 
    \end{tabular}
  \caption{Representation results related to this paper, where monotonicity is always assumed; definitions of the properties are in Sections \ref{sec:pre} and \ref{sec:new3}. CA stands for cash additivity and CS stands for cash subadditivity. 
   \label{tab:1}}
  \end{center}
\end{table}

Furthermore, we obtain several results  on the risk measure $\Lambda\VaR$ proposed by \cite{FMP14}, including a  new representation result (Theorem \ref{prop:lambdarep}). In particular,  the class of $\Lambda\VaR$ serves as a natural example of quasi-star-shaped, quasi-normalized and cash-subadditive risk measures, which are not star-shaped, normalized, or cash additive. 

Risk measures without cash additivity have received increasing attention in the recent literature due to their technical generality and intimate connection to decision analysis, risk transforms, portfolio optimization, and stochastic interest rates; many references and examples were mentioned in the introduction and throughout the paper. 
Results in this paper serve as a building block for future studies on cash subadditivity and the new properties of quasi-star-shapedness and quasi-normalization, for which many questions and applications remain to be explored. 


\subsection*{Acknowledgments}
We thank Fabio Maccheroni, Giulio Principi,  Peter Wakker, Qinyu Wu, an Editor, an Associated Editor, and two anonymous referees for helpful comments on the paper. J.~Xia would like to acknowledge the financial support of the National Key R\&D Program of China (grant 2018YFA0703900) and the National Natural Science Foundation of China (grants 12071146, 12431017, 12471447). 
X.~Han, Q.~Wang and R.~Wang acknowledge  financial support from the Natural Sciences and Engineering Research Council of Canada (RGPIN-2024-03728, CRC-2022-00141). X.~Han is also supported by  the National Natural Science Foundation of China (grants No.~12301604, 12371471, 12471449). Q.~Wang is also supported by the Society of Actuaries through the James C.~Hickman Scholar Doctoral Stipend.




\appendix



\section{Additional   results and technical discussions}\label{app:A}
This appendix includes a few additional technical results, examples and discussions of the representation results of cash-subadditive risk measures, which are not used in the main text of the paper. Some of them may be of independent interest.

\subsection{Connection to a representation of monetary risk measures}\label{App:1}

{Theorem \ref{thm:1} is more general than the result of \cite{JXZ21} for monetary risk measures, which says that any monetary risk measure can be written as the infimum of some convex monetary risk measures.   Indeed, the proof of Theorem \ref{thm:1} does not depend on, but leads to Theorem 3.1 of \cite{JXZ21} as a special case.} 

The proposition below illustrates how Lemma \ref{lem:represent} (the key lemma proving Theorem \ref{thm:1}) can be used to show that any monetary risk measure is the minimum of some convex risk measures.

\begin{proposition}\label{prop:JXZ}
A functional $\rho:\X\to\R$ is a monetary risk measure if and only if
$$\rho(X)=\min_{Z\in\mathcal{A}_\rho}\esssup(X-Z), ~~\text{for all }X\in \X,$$
where $\mathcal{A}_\rho$ is the acceptance set of $\rho$ given by $\mathcal{A}_\rho=\{Z\in \X\mid \rho(Z)\le 0\}$.
\end{proposition}
\begin{proof}
The ``if" part is straightforward. We prove the ``only if" part. For all $X\in \X$, by Lemma \ref{lem:represent} and cash additivity of $\rho$, we have
$$\rho(X)=\min_{Z\in \X}\rho(Z+\esssup(X-Z))=\min_{Z\in \X}\left\{\rho(Z)+\esssup(X-Z)\right\}.$$
By taking $Z_0=X-\rho(X)$, we have $\rho(Z_0)+\esssup(X-Z_0)=\rho(X)$, where the minimum is obtained. Define $\mathcal{A}_\rho^0=\{Z\in\X\mid\rho(Z)=0\}$. We have $Z_0\in\mathcal{A}_\rho^0$ and thus
\begin{align*}
\rho(X)=\min_{Z\in \mathcal{A}_\rho^0}(\rho(Z)+\esssup(X-Z))=\min_{Z\in \mathcal{A}_\rho^0}\esssup(X-Z)\ge\min_{Z\in\mathcal{A}_\rho}\esssup(X-Z).
\end{align*}
On the other hand, since $\rho(Z)\le 0$ for all $Z\in\mathcal{A}_\rho$, we have
\begin{align*}
\rho(X)=\min_{Z\in\mathcal{A}_\rho}\left\{\rho(Z)+\esssup(X-Z)\right\}\le\min_{Z\in\mathcal{A}_\rho}\esssup(X-Z).
\end{align*}
Therefore, we have $\rho(X)=\min_{Z\in\mathcal{A}_\rho}\esssup(X-Z)$.
\end{proof}

\subsection{Comonotonic quasi-convexity} \label{sec:uni}

{Let $(\Omega, \mathcal{F}, P)$ be a nonatomic probability space.} Since  law-invariant, quasi-convex and {{cash-subadditive}} risk measures are SSD-consistent (Lemma \ref{lem:SSD}), 
a general law-invariant cash-subadditive risk measure (such as VaR in Section \ref{sec:pre}) does not  admit a representation via the lower envelope of a family of law-invariant, quasi-convex and {cash-subadditive} risk measures.
 One remaining question is whether a
 law-invariant cash-subadditive risk measure can be represented as the infimum of a set of law-invariant cash-subadditive risk measures with some other properties. For such a representation, 
 we need comonotonic quasi-convexity.
 \begin{itemize}
  \item[] \emph{Comonotonic quasi-convexity}:   $\rho(\lambda X+(1-\lambda) Y) \leq \max \{\rho(X), \rho(Y)\}$ for all comonotonic  $(X, Y) \in \X^2$  and $\lambda \in[0,1]$. 
 \end{itemize}
 The property of comonotonic quasi-convexity appeared in various contexts; e.g., \cite{X13}, \cite{TL15} and \cite{LW19}. 
 Before showing the representation result, we first give the following equivalence result demonstrating the relations among several properties of $\rho$, similarly to Lemma \ref{lem:SSD}.
\begin{lemma}\label{lem:comcx}
If $\rho:\X\to\R$ is a cash-subadditive risk measure, then $\rho$ is law invariant and quasi-convex if and only if $\rho$ is SSD-consistent and comonotonic quasi-convex.
\end{lemma}
\begin{proof}
The ``only if" part follows directly from Lemma \ref{lem:SSD}. We prove the ``if" part. Suppose that $\rho$ is SSD-consistent and comonotonic quasi-convex. It is clear that $\rho$ is law invariant by taking $X\laweq Y$ and observing $X\succeq_2 Y$ and $Y\succeq_2 X$. For all $X,Y\in\X$, take $X^c,Y^c\in\X$ such that $(X^c,Y^c)$ is comonotonic, $X^c\laweq X$, and $Y^c\laweq Y$. {Again, by Theorem 3.5 of \cite{R13}, we have  $\lambda X^c+(1-\lambda)Y^c\succeq_2 \lambda X+(1-\lambda)Y$ for all $\lambda\in[0,1]$.}  Hence, we have
$$\rho(\lambda X+(1-\lambda)Y)\le \rho(\lambda X^c+(1-\lambda)Y^c)\le \max\{\rho(X^c),\rho(Y^c)\}=\max\{\rho(X),\rho(Y)\},$$
which indicates that $\rho$ is quasi-convex.
\end{proof}
 
With the extra requirement of  comonotonic quasi-convexity,
we obtain a unifying umbrella for the representation of cash-subadditive risk measures with various properties. 
This result is parallel to the result of \cite{JXZ21} on monetary risk measures, where comonotonic convexity \citep{SY06, SY09} is in place of our comonotonic quasi-convexity.

\begin{proposition}\label{prop:CoM} For a functional $\rho: \X \rightarrow \mathbb{R},$ we have the following statements.
 \begin{enumerate}[(i)]
\item $\rho$ is a cash-subadditive  risk measure if and only if it is the lower envelope of a family of comonotonic quasi-convex and cash-subadditive risk measures. 

\item  $\rho$ is a law-invariant cash-subadditive  risk measure if and only if it is the lower envelope of a family of law-invariant, comonotonic quasi-convex and cash-subadditive  risk measures. 
\end{enumerate}
The equivalence (ii) holds true if ``law-invariant" is replaced by ``normalized and quasi-star-shaped" or ``SSD-consistent".
\end{proposition}

\begin{proof}
Note that each of law invariance, normalization, quasi-star-shapedness, SSD-consistency and cash subadditivity is preserved under taking an infimum, and hence the ``if" parts in all statements are obvious. Since comonotonic quasi-convexity  is weaker than  quasi-convexity,  
the representations (``only if") in Theorems \ref{thm:1}, \ref{thm:PH} and \ref{thm:SSD} hold   true by replacing quasi-convexity with comonotonic quasi-convexity. 
This, together with Lemma \ref{lem:SSD}, gives the ``only if" parts except for the case of law-invariant cash-subadditive risk measures in (ii).  Below we show this part.

Assume $\rho$ is a law-invariant cash-subadditive  risk measure.  According to  Proposition \ref{thm:LI} below, for all $X\in\X$, we have $\rho(X)=\min_{Z\in \X}\phi_Z(X)$ in which 
 $$\phi_Z(X)=\rho\left(Z+\sup_{t\in(0,1)}(\VaR_t(X)-\VaR_t(Z))\right).$$
It is clear that $\phi_Z$ is monotone, cash subadditive and law invariant. We prove that $\phi_Z$ is comonotonic quasi-convex  by the similar way to Theorems \ref{thm:1} and \ref{thm:SSD}. Assume that   $(X_1, X_2)\in \X^2$ is comonotonic and $\phi_{Z}\left(X_{i}\right) \leq \alpha$ for $i=1,2$.  For all $\varepsilon>0$ and $i=1,2,$ there exists some $m_i \in \mathbb{R}$ such that $Z+m_i \succeq_1 X_{i}$ and $\rho\left(Z+m_i\right) \leq \phi_{Z}\left(X_{i}\right)+\varepsilon \leq \alpha+\varepsilon $. 
For all $\lambda\in[0,1]$, comonotonic additivity of $\VaR_t$ yields that $$\VaR_t(\lambda X_{1}+(1-\lambda) X_{2})=\lambda\VaR_t(X_1)+(1-\lambda)\VaR_t(X_2)\le\VaR_t(Z+\lambda m_1+(1-\lambda)m_2),$$
for all $t\in(0,1)$.
We thus have
$Z+m\succeq_1 \lambda X_{1}+(1-\lambda) X_{2}$.
It follows that
$$
\phi_{Z}\left(\lambda X_{1}+(1-\lambda) X_{2}\right) \leq \rho(Z+\lambda m_1+(1-\lambda)m_2)\leq \rho(Z+\max\{m_1,m_2\}) \leq \alpha+\varepsilon.
$$ 
Since $\varepsilon$ is arbitrary, $\phi_{Z}\left(\lambda X_{1}+(1-\lambda) X_{2}\right) \leq \alpha .$ Therefore, $\phi_{Z}$ is comonotonic quasi-convex.
\end{proof}

\subsection{Law-invariant cash-subadditive risk measures and VaR}
\label{sec:law}
{Let $(\Omega, \mathcal{F}, P)$ be a nonatomic probability space} in this section. We first connect law-invariant cash-subadditive risk measures to VaR defined in Section \ref{sec:pre}. It is well known that the class of VaR characterizes FSD via
$$X \succeq_{1} Y\iff \VaR_t(X)\ge\VaR_t(Y) \text{ for all }t\in(0,1).$$  
\begin{proposition}\label{thm:LI} 
If $\rho:\X\to\R$ is a risk measure, then $\rho$ is law invariant and cash subadditive if and only if it satisfies
$$\rho(X)=\min_{g\in\mathcal{G}_X}\sup_{t\in(0,1)}\left\{\VaR_t(X)-g(t)\right\},~~\text{for all }X\in\X,$$
where $\mathcal{G}_X$ is a set of measurable functions from $(0,1)$ to $(-\infty,\infty]$ for all $X\in\X$, with $\mathcal{G}_{X_1}\subseteq\mathcal{G}_{X_2}$ for all $X_1,X_2\in \X$ such that $X_2\succeq_1 X_1$.
Moreover,  the set $\mathcal{G}_X$ can be chosen as
$$\mathcal{G}_X=\{g:(0,1)\to(-\infty,\infty],~t\mapsto \VaR_t(Z)-\rho(Z)\mid Z\in\X,~X\succeq_1 Z\},~~\text{for all }X\in\X.$$

\end{proposition} 

\begin{proof}

``$\Rightarrow$": Suppose that $\rho$ is a law-invariant cash-subadditive risk measure. For all $X\in\X$ and $Z\in \X$, define the risk measure
$$
\phi_{Z}(X)=\inf \{\rho(Z+m) \mid m \in \mathbb{R},~ Z+m\succeq_{1}  X\}.
$$
For all $m\in\R$ such that $Z+m\succeq_1 X$, since any law-invariant risk measure is FSD-consistent \citep[e.g.,][Remark 4.58]{FS16}, we have $\rho(Z+m)\ge \rho(X)$. It follows that $\phi_{Z}(X)\ge\rho(X)$ for all $Z\in \X$. Noting that $\phi_X(X)=\rho(X)$, we have $\rho(X)=\min_{Z\in \X}\phi_Z(X)$.
By definition of $\phi_Z$, we have
\begin{equation*}
\begin{aligned}
\phi_{Z}(X)&=\inf \left\{\rho(Z+m) \mid m \in \mathbb{R},~ \VaR_t(Z)+m\ge  \VaR_t(X)~\text{for all }t\in(0,1)\right\}\\
&=\rho\left(Z+\sup_{t\in(0,1)}(\VaR_t(X)-\VaR_t(Z))\right).
\end{aligned}
\end{equation*}
 Further, we have
 \begin{equation*}
 \begin{aligned}
\rho(X)&=\min_{Z\in \X,~X\succeq_1 Z}\sup_{t\in(0,1)}\rho(Z+\VaR_t(X)-\VaR_t(Z))\\
&\le \min_{Z\in \X,~X\succeq_1 Z}\sup_{t\in(0,1)}\left\{\VaR_t(X)-\VaR_t(Z)+\rho(Z)\right\}\le \rho(X).
\end{aligned}
\end{equation*}
It follows that
$$\rho(X)=\min_{Z\in \X,~X\succeq_1 Z}\sup_{t\in(0,1)}\left\{\VaR_t(X)-g_Z(t)\right\},$$
where $g_Z(t)=\VaR_t(Z)-\rho(Z)$ for all $t\in(0,1)$. Therefore, $\{g_Z\mid Z\in \X,~X\succeq_1 Z\}$ is a desired family of measurable functions on $(0,1)$.
 
 ``$\Leftarrow$": We first show that $\rho$ is cash subadditive. Indeed, for all $X\in\X$ and $m\ge 0$, we have $X+m\succeq_1 X$. Hence, $\mathcal{G}_X\subseteq\mathcal{G}_{X+m}$ and 
 \begin{align*}
 \rho(X+m)&=\min_{g\in\mathcal{G}_{X+m}}\sup_{t\in(0,1)}\left\{\VaR_t(X)-g(t)\right\}+m\\
 &\le \min_{g\in\mathcal{G}_X}\sup_{t\in(0,1)}\left\{\VaR_t(X)-g(t)\right\}+m=\rho(X)+m.
 \end{align*}
 To show law invariance of $\rho$, for all $X,Y\in\X$ such that $X\laweq Y$, we have $X\succeq_1 Y$ and $Y\succeq_1 X$. It follows that $\mathcal{G}_X=\mathcal{G}_Y$ and thus $\rho$ is law invariant.
%
\qedhere
\end{proof}


 \begin{remark}
 \label{rem:LI}
Although the functional
$$
\phi_{Z}(X)=\inf \{\rho(Z+m) \mid m \in \mathbb{R},~ Z+m\succeq_{1}  X\},~~X\in\X,
$$
defined in the proof of Proposition \ref{thm:LI} is monotone, cash subadditive and law invariant, $\phi_Z$ is not quasi-convex.  This is because $\VaR$ does not satisfy quasi-convexity. 
\end{remark}


\subsection{Certainty equivalents of $\alpha$-maxmin expected utility}

{We assume $(\Omega, \mathcal{F}, P)$ be a nonatomic probability space.} The  rank-dependent expected utility (RDEU)  of \cite{Q82} is a popular behavioral decision model specified by the preference functional
$$\int_\Omega \ell(X)\d T\circ P,~~~X\in \X,$$
where $\ell:\R\to\R$ is a strictly increasing and convex loss function (positive random variables represent losses), $T:[0,1]\to[0,1]$ is a probability distortion function, {and the integral with respect to $T\circ P$ is a Choquet integral (\cite{C54}, \cite{S86,S89})}.  We consider the  choice  of  $T $ given by $T=\alpha T_1+(1-\alpha)T_2$ where $T_1$ (resp.~$T_2$) are increasing, differentiable and convex (resp.~concave) probability distortion functions with $T_1(0)=T_2(0)=0$ and $T_1(1)=T_2(1)=1$.
{This corresponds to the well known $\alpha$-maximin model of \cite{M02} and \cite{GMM04} (see Example \ref{ex:alphaMEU}),
with the interpretation of balancing between optimistic and pessimistic views on ambiguity.}
Following \cite{CD03}, for an increasing, differentiable and convex distortion function $h:[0,1]\to[0,1]$ with $h(0)=0$ and $h(1)=1$, define the core  of $h\circ P$ by 
$$\mathrm{core}(h\circ P)=\{Q\in\mathcal{M}\mid Q(A)\ge h(P(A)) \text{ for all }A\in\mathcal{F}\}.$$
Continuity of $h$ guarantees that any element in the core of $h\circ P$ is a probability measure absolutely continuous with respect to $P$, which may be identified with its density with respect to $P$.
 Thus, we have 
\begin{equation}\label{eq:core}
\int_\Omega \ell(X)\d T\circ P=\alpha\min_{Q_1\in\mathrm{core}(T_1\circ P)}\E_{Q_1}[\ell(X)]+(1-\alpha)\max_{Q_2\in\mathrm{core}(\widehat{T}_2\circ P)}\E_{Q_2}[\ell(X)],
\end{equation}
where $\widehat{T}_2: x\mapsto1-T_2(1-x)$.
  The certainty equivalent of the RDEU with an ambiguous discount factor $\lambda$ is given by
\begin{equation}\label{eq:CE}
\begin{aligned}
\rho(X)&=\sup_{\lambda\in I}~\ell^{-1}\left(\int_\Omega \ell(\lambda X)\d T\circ P\right)\\
&=\min_{Q_1\in\mathrm{core}(T_1\circ P)}\max_{Q_2\in\mathrm{core}(\widehat{T}_2\circ P)}\sup_{\lambda\in I}~\ell^{-1}\left(\alpha\E_{Q_1}[\ell(\lambda X)]+(1-\alpha)\E_{Q_2}[\ell(\lambda X)]\right),
\end{aligned}
\end{equation}
where  $I \subseteq[0,1]$ is the ambiguity set. 
{For technical tractability, we assume that the discount factor is deterministic here.}
 It is clear that if we take the loss function to be $\ell:x\mapsto\mathrm{e}^{\gamma X}$ for $\gamma>0$, then $\rho$ is a cash-subadditive risk measure, while $\rho$ becomes a monetary risk measure without ambiguity of the discount factor $\lambda$.

Note that for all $\lambda\in I$, $Q_1\in\mathrm{core}(T_1\circ P)$ and $Q_2\in\mathrm{core}(\widehat{T}_2\circ P)$, the mapping $$X\mapsto\ell^{-1}\left(\alpha\E_{Q_1}[\ell(\lambda X)]+(1-\alpha)\E_{Q_2}[\ell(\lambda X)]\right)$$ is quasi-convex and upper semi-continuous. Proposition 5.3 of \cite{CMMM11} showed an explicit representation of the certainty equivalent of the expected loss given by $\ell^{-1}(\E_P[\ell(\cdot)])$. In the proposition below, we show the representation result of a more general $\rho$ in a similar sense. Define $\bar{\ell}:[-\infty,\infty]\to[-\infty,\infty]$ as the extended-valued function with inverse function given by
$$\bar{\ell}^{-1}(x)=\left\{\begin{array}{ll}
\ell^{-1}(x), & x\in(\inf_{t\in\R}\ell(t),\infty),\\
-\infty,& x\in[-\infty,\inf_{t\in\R}\ell(t)],\\
\infty, & x = \infty.
\end{array}\right.$$
Let $\ell^*:[-\infty,\infty]\to[-\infty,\infty]$ be the conjugate function of $\bar{\ell}$ given by
$$\ell^*(x)=\sup_{y\in[-\infty,\infty]}\{xy-\bar{\ell}(y)\},~~x\in[-\infty,\infty].$$

\begin{proposition}\label{prop:dual}
Let $\widetilde{Q}=\alpha Q_1+(1-\alpha)Q_2$ for  $Q_1\in\mathrm{core}(T_1\circ P)$ and $Q_2\in\mathrm{core}(\widehat{T}_2\circ P)$.
For $X\in\X$, the risk measure $\rho$ in \eqref{eq:CE} adopts the following representation:
$$\rho(X)=\min_{Q_1\in\mathrm{core}(T_1\circ P)}\max_{Q_2\in\mathrm{core}(\widehat{T}_2\circ P)}\sup_{\lambda\in I}\max_{Q\in{\mM}}~R\left(\E_Q[X],\lambda,Q, \widetilde Q\right),$$
where 
\begin{align}
    \label{eq:r1-CE}
{ R\left(t,\lambda,Q,\widetilde Q\right )}=\ell^{-1}\left(\max _{x \geq 0}\left[\lambda x t-\mathbb{E}_{\widetilde{Q}}\left(\ell^{*}\left(x \frac{\d Q}{\d \widetilde{Q}}\right)\right)\right]\right),~~{(t,\lambda, Q, \widetilde Q) \in \mathbb{R} \times I \times \mathcal{M}\times \mathcal M.}\end{align}
\end{proposition}
\begin{proof}
{For   $X\in\X$, $\lambda\in I$, $Q_1\in\mathrm{core}(T_1\circ P)$ and $Q_2\in\mathrm{core}(\widehat{T}_2\circ P)$,
we have 
$$\ell^{-1}\left(\alpha\E_{Q_1}[\ell(\lambda X)]+(1-\alpha)\E_{Q_2}[\ell(\lambda X)]\right)=\ell^{-1}\left(\E_{\widetilde{Q}}[\ell(\lambda X)]\right).$$ 
 Proposition 5.3 of \cite{CMMM11} gives
$$\ell^{-1}\left(\E_{\widetilde{Q}}[\ell(\lambda X)]\right) =\max_{Q\in{\mM}}{R(\E_{{Q}}[X],\lambda,Q,\widetilde Q)},$$
where $R(t,\lambda,Q,\widetilde Q)$ is given by \eqref{eq:r1-CE}. 
This and \eqref{eq:CE} yield the desired representation.}
\end{proof}

{Proposition \ref{prop:dual} gives, for the $\alpha$-maxmin risk measure in this section, the explicit representation in  the form of Theorem \ref{thm:1}.}

\subsection{A counter-example}\label{app:A5}
\begin{example}[$\Lambda\VaR$ is not quasi-star-shaped]\label{exm:non-qqs}For $0<\alpha<1/2<\beta< 1$, consider the  increasing function  $\Lambda(x)=\alpha\id_{\{x\le1/2\}}+\beta\id_{\{x> 1/2\}},~x\in\R.$
For $t=7/4$, $\lambda=1/8$, and a Bernoulli random loss $X$ given by
$P(X=2)=P(X=0)=1/2$, we have  $P(\lambda X+(1-\lambda)t=57/32)=P(\lambda X+(1-\lambda)t=49/32)=1/2.$  Hence,  $\Lambda\VaR(X)=\VaR_\alpha(X)=0,~\Lambda\VaR(\lambda X+(1-\lambda)t)=\VaR_\beta(\lambda X+(1-\lambda)t)=57/32,$ 
and $\Lambda\VaR(t)=7/4$. It follows that
 $\Lambda\VaR(\lambda X+(1-\lambda)t)>\max\{\Lambda\VaR(X),\Lambda\VaR(t)\}$
and $\Lambda\VaR$ is not quasi-star-shaped.
\end{example}

 \end{document}